\documentclass[12pt]{article}
\usepackage{color, amsmath, amsthm}
 
\usepackage{graphicx}
\usepackage{amsfonts}
\usepackage{amssymb,amsmath,amscd}
\usepackage{fancyhdr}
\usepackage[top=2.75cm, bottom=2.75cm, left=2.75cm, right=2.75cm]{geometry} 
\usepackage{slashed}
\usepackage{hyperref}
\usepackage[toc,page]{appendix}
\numberwithin{equation}{section}
\hypersetup{
    pdftitle={Titlel},    
    pdfauthor={Yiwen Pan},     
    pdfnewwindow=true,      
    colorlinks=true,       
    linkcolor=black,          
    citecolor= black,        
    filecolor= black,      
    urlcolor= black           
}

\begin{document}

\title{Note on a Cohomological Theory of Contact-Instanton and Invariants of Contact Structures}
\date{}
\author{Yiwen Pan\footnote{Email address:  \href{mailto:yiwen.pan@stonybrook.edu}{yiwen.pan@stonybrook.edu}}
\\ \textit{Department of Physics and Astronomy, Stony Brook}\\
    \textit{Institute Address}}
\maketitle
\begin{abstract}
	\noindent In the localization of 5-dimensional $\mathcal{N} = 1$ super-Yang-Mills, contact-instantons arise as non-perturbative contributions. In this note, we revisit such configurations and discuss their generalizations. We propose for contact-instantons a cohomological theory whose BRST observables are invariants of the background contact geometry. To make the formalism more concrete, we study the moduli problem of contact-instanton, and we find that it is closely related to the eqiuivariant index of a canonical Dirac-Kohn operator associated to the geometry. An integral formula is given when the geometry is K-contact. We also discuss the relation to 5d $\mathcal{N} = 1$ super-Yang-Mills, and by studying a contact-instanton solution canonical to the background geometry, we discuss a possible connection between $\mathcal{N} = 1$ theory and contact homology. We also uplift the 5d theory a 6d cohomological theory which localizes to Donaldson-Uhlenbeck-Yau instantons when placed on special geometry.
\end{abstract}

\thispagestyle{fancy}



\newpage
\tableofcontents

\section{Introduction}

	In \cite{Witten}, a quantum field theory is proposed whose expectation values of observables computes invariants proposed by Donaldson. One of the key ingredient of the theory is the 4-dimension instanton moduli space, which is sensitive to the smooth structure of the underlying manifold.
	
	It is reasonable to believe that instantons, or more generally, connections whose curvatures satisfy certain involutive linear algebraic equations, can encode interesting information about the geometry of the base manifold. There are previous efforts trying to extend Witten's work to higher dimensions, by defining corresponding instanton-like objects and follow the cohomological field theory paradigm. Note that the 4-dimension instanton defined by involution $*$, cannot be directly applied to other dimensions; already discussed in early literature \cite{Corrigan:1983}, additional geometry structures are often required\footnote{Actually, in \cite{Corrigan:1983}, the contact instanton equation was proposed, although in that context it is not an interesting equation.}. In particular, previous works mainly focus on 6, 7, 8-dimensional manifolds, equipped with Calabi-Yau, $G_2$ and ${\rm Spin}(7)$ geometries \cite{Corrigan:1983}\cite{B.S.Acharya:aa}\cite{Donaldson:2009aa}. There is relatively a large gap in understanding similar phenomena in 5d.
	
	Recently there are new interests in five-dimensional gauge theories in the last few years. In particular, the 5d maximal ($\mathcal{N} = 2$) supersymmetric Yang-Mills theory is believed to hold key or even all the information of the 6d (2,0)-theory. In particular, the solitonic solutions (instantons on the 4-dimensional space slice) of the  Yang-Mills theory correspond to the K-K modes of the $S^1$-compatification from 6d\cite{Lambert:2010aa}. On the other hand, in \cite{Kallen:2012aa}, the partition function of $\mathcal{N} = 1$ super-Yang-Mills theory on K-contact manifolds is studied by localization method, and in particular, the localization locus are the configurations of gauge fields $A$ satisfying
	\begin{equation}
		\iota_R F_A = 0, \;\;\;\; \pi_H^- F_A = 0,
		\label{eq1}
	\end{equation}
which are collectively called contact-instantons. In special cases, these contact-instanton reduces to usual 4-dimension self-dual instantons. Intuitively, a contact-instanton consists of infinitely many 4d self-dual instantons distributed on the 5-dimensional K-contact manifold. At the same time, there are parallel mathematical works on the twistor formulation of contact-instanton \cite{Wolf:2012aa}, and the relation with Killing spinors\footnote{In \cite{Harland:2011aa} there are 3 different definition of ``instanton", and the third definition is the one that agrees with \cite{Kallen:2012aa}.} \cite{Harland:2011aa}.

	It is then natural to wonder if such new notion of instanton gives any information of the geometry on which it is defined. Notice that the equation (\ref{eq1}) can be defined for any $\kappa$ and $R$ with unit norm. However, by relating to Yang-Mills equation, we see that it is better to restrict $(\kappa, R, g)$ to be contact.
	
	As we will see, on a contact geometry, one can define a generalized notion of contact-instanton as configurations satisfying
	\begin{equation}
		\pi _V {F_A} = 0,\;\;\pi _H^ + {F_A} = \lambda \phi d\kappa, \;\;d_A \phi  = 0,\;\;\lambda \in \mathbb{R}.
		\label{contact-instanton-intro}
	\end{equation}
We claim that they contain interesting information about contact structures. Following the formalism of cohomological field theory, we construct a theory with BRST charge $Q$ that localizes onto (\ref{contact-instanton-intro}), and its deformation under homotopy of background geometry is $Q$-exact. Formally, the expectation values of BRST observables then gives invariants of contact structures.

	On K-contact manifolds, the dimension of moduli space of (\ref{contact-instanton-intro}) can be computed in certain gauge choice. It turns out that it is related to the equivariant index of some Dirac-like operator on the canonical ${\rm Spin}^\mathbb{C}$ bundle associated with contact metric structures\cite{Fitzpatrick:2007aa}. We also discuss a vanishing theorem, after which we gives an integral formula to compute the dimension of the moduli space.
	
	The theory is closely related with $\mathcal{N} = 1$ super-Yang-Mills. We show that the twisted $\mathcal{N} = 1$ theory gives information about K-contact structures. In particular, on any Sasaki-Einstein manifold there is a canonical solution to (\ref{contact-instanton-intro}), to which the twisted $\mathcal{N} = 1$ theory localizes. The corresponding Wilson loops $W(\gamma)$ coincide with the linearized return map $\Psi_\gamma$ in contact homology. We also uplift the theory to 6d and discuss the relation with Donaldson-Uhlenbeck-Yau equation on Calabi-Yau 3-fold.
	
	\vspace{10pt}
	This note is organized as follows.
	
	\underline{Section 2}: we review the notion of self-duality on a five dimensional Riemannian manifold $M$ equipped with a nowhere-vanishing 1-form. We then define contact-instanton on $M$ and discuss its relation with Yang-Mills equation, which urges us to focus on contact geometry.
	
	\underline{Section 3}: we summarize the definitions and important properties of contact geometry, and list a few useful formula that will be used subsequently. In the appendix \ref{Appendix-B} we provide more detail on the subject.
	
	\underline{Section 4}: we propose a cohomological theory with BRST charge $Q$, along with its BRST observables. We then discuss the invariance under homotopy of underlying contact structure, proving that they are invariants of contact structures.
	
	\underline{Section 5}: we study the dimension of moduli space $\mathcal{M}$ of contact-instanton. We reformulate the deformation problem in specific gauge and focus on K-contact manifolds. By straight forward computations, we relate the dimension $\dim \mathcal{M}$ with the equivariant index of a canonical Dirac-like operator on $S$, the canonical ${\rm Spin}^\mathbb{C}$ spinor bundle. An integral formula is shown by referring to known mathematical literature.

	\underline{Section 6}: we compare the cohomological theory with twisted $\mathcal{N} =1$ super-Yang-Mills theory. We show a possible relation between the Wilson loops $\Psi_\gamma$ along closed Reeb orbits $\gamma$, and the linearize return map in contact homology. We also uplift the cohomological theory to 6d, and show that the localization locus of the 6d theory are solutions of Donaldson-Uhlenbeck-Yau equation on Calabi-Yau 3-folds.
	
	In the appendices, we first review our notations and some basic differential geometry. Then we provide a pedagogical review of contact geometry, adding in details that are neglected in section 3. In appendix [\ref{Appendix-C}], we review in detail the canonical ${\rm Spin}^\mathbb{C}$ structure on a contact metric manifold. We discuss generalized Tanaka-Webster connection and its relation to a Dirac-Kohn operator on the canonical spinor bundle, which plays important role in section 5.
	
	{\em Note}: before the submission of this note, a mathematics paper is published on arxiv \href{http://arxiv.org/abs/1401.5140}{\em{\underline {Moduli Spaces of Contact Instantons}}}\cite{Baraglia:2014aa}, which focuses on the contact-instanton moduli space, discusses conditions for smoothness and geometry .

\section{Contact-Instanton}

	\subsection{Self-duality in $d = 5$}

	Let $M$ be a 5-dimensional smooth manifold. Since the Euler characteristics $\chi(M) = 0$, the bundles $TM$ and $T^*M$ both have nowhere-vanishing sections. Let us choose one such $\kappa \in \Gamma(T^*M)$. We also select $R \in \Gamma(TM)$ such that
	\begin{equation}
		{\iota _R}\kappa  = 1.
	\end{equation}
With these two quantities, one can decompose any 2-form $\omega$
	\begin{equation}
		\omega  = {\omega _V} + {\omega _H} = \kappa  \wedge {\iota _R}\omega  + {\iota _R}\left( {\kappa  \wedge \omega } \right),
	\end{equation}
where $V$ stands for ``vertical"" and H ``horizontal". Accordingly, the space of 2-forms is decomposed
	\begin{equation}
		{\Omega ^2}\left( M \right) = \Omega _V^2\left( M \right) \oplus \Omega _H^2\left( M \right).
	\end{equation}
	
	We continue to choose a Riemann metric $g$ such that $\kappa$ has unit norm:
	\begin{equation}
		g\left( {R, \cdot } \right) = \kappa \left(  \cdot  \right).
	\end{equation}

	With the data $(\kappa, R, g)$, one can further decompose any horizontal forms $\Omega_H$
	\begin{equation}
		{\omega _H} = \omega _H^ +  + \omega _H^ -  = \frac{1}{2}\left( {1 + {\iota _R}*} \right){\omega _H} + \frac{1}{2}\left( {1 - {\iota _R}*} \right){\omega _H},
	\end{equation}
and accordingly the decomposition $\Omega _H^2\left( M \right) = \Omega _H^ + \left( M \right) \oplus \Omega _H^ - \left( M \right)$. Note that acting on any $p$-form $\omega_p$ one has relation
	\begin{equation}
		{\iota _R}*{\omega _p} = {\left( { - 1} \right)^p}*\left( {\kappa  \wedge {\omega _p}} \right),
	\end{equation}
and therefore on horizontal 2-forms
	\begin{equation}
		{\left( {{\iota _R}*} \right)^2} = 1,\;\;\;\; {\rm on} \; \Omega_H^2(M).
	\end{equation}
and we call $\Omega_H^+$ self-dual while $\Omega_H^-$ anti-self-dual 2-forms. Note that the wedge product between the two types of forms vanishes:
	\begin{equation}
		\begin{gathered}
		  {\omega _ + } \wedge {\omega _ - } =  - {\omega _ + } \wedge {\iota _R}*{\omega _ - } =  - {\iota _R}\left( {{\omega _ + } \wedge *{\omega _ - }} \right) =  - {\iota _R}\left( {{\omega _ - } \wedge *{\omega _ + }} \right) =  - {\omega _ - } \wedge {\iota _R}*{\omega _ + } \hfill \\
		  \;\;\;\;\;\;\;\;\;\;\;\;\;\; =  - {\omega _ + } \wedge {\omega _ - } \hfill .
		  \label{vanishing-product}
		\end{gathered} 
	\end{equation}
	
	Let us denote the projection onto the various spaces for later convince:
	\begin{equation}
		{\pi _H} \equiv {\iota _R}\circ (\kappa  \wedge) ,\;\;{\pi _V} \equiv \kappa  \wedge {\iota _R},\;\;\pi _H^ \pm  = \frac{1}{2}\left( {{\pi _H} \pm {\iota _R}*} \right),\;\; \pi^\pm \equiv \frac{1}{2} (1\pm \iota_R *).
		\label{projections}
	\end{equation}
Note that the all the above operators but $\pi^\pm$ square to themselves, while
	\begin{equation}
		{\left( {{\pi ^ \pm }} \right)^2} = \frac{1}{4}{\pi _V}.
	\end{equation}
To summarize, any 2-form $\omega$ can be decomposed into three parts:
	\begin{equation}
		\begin{gathered}
  		\omega  = {\pi _V}\omega  + \pi _H^ + \omega  + \pi _H^ - \omega  \equiv {\omega _V} + \omega _H^ +  + \omega _H^ -  \hfill \\
  		\;\;\;\;\;\;\;\;\;\;\;\;\;\;\;\;\;\;\;\;\;\;\;\;\;\;\;\;\;\;\;\;\;\;\;\;\;= \left( {\frac{1}{2}{\omega _V} + \omega _H^ + } \right) + \left( {\frac{1}{2}{\omega _V} + \omega _H^ - } \right) \hfill \\
  		\;\;\;\;\;\;\;\;\;\;\;\;\;\;\;\;\;\;\;\;\;\;\;\;\;\;\;\;\;\;\;\;\;\;\;\;\; = {\pi ^ + }\omega  + {\pi ^ - }\omega.  \hfill \\ 
		\end{gathered} 
	\end{equation}
	
	Finally, let us point out that the decomposition is orthogonal with respect to the inner product
	\begin{equation}
		\left( {\omega _1^p,\omega _2^p} \right)  \equiv \int_M {\omega _1^p \wedge *\omega _2^p} .
	\end{equation}
Moreover, all the operators are self-adjoint, namely ${\left( {\pi _H^ \pm } \right)^*} = \pi _H^ \pm $, ${\pi _V^*} = {\pi _V}$:
	\begin{equation}
		\left( {\omega _1^p,\pi _H^ \pm \omega _2^p} \right)  = \left( {\pi _H^ \pm \omega _1^p,\omega _2^p} \right) ,\;\;\;\;\left( {\omega _1^p,{\pi _V}\omega _2^p} \right) = \left\langle {{\pi _V}\omega _1^p,\omega _2^p} \right) .
	\end{equation}
and in particular, $\iota_R *$ is also self-adjoint.
	
	\subsection{Contact Instanton and Yang-Mills equation}
	
	Let  $G$ be a semi-simple Lie group with Lie algebra $\mathfrak{g}$. Denote $P_G$ as a principal $G$-bundle over $M$, and ${\rm ad}P_G$ as the associated adjoint vector bundle. Denote $\mathcal{A}$ the space of connection of ${\rm ad}P_G$, and $\mathcal{G}$ the space of gauge transformation. The section of ${\rm ad}P_G$ is denoted as $\Omega^0(M,\mathfrak{g})$, and similarly the space of $p$-forms valued in ${\rm ad}P_G$ is denoted as $\Omega^p(M,\mathfrak{g})$. We want $A$ to be hermitian, and we denote
	\begin{equation}
		{d_A}\phi  = d\phi  - i\left[ {A,\phi } \right],\;\;\phi \in \Omega^0(M, \mathfrak{g}).
	\end{equation}
	
	Using the projection operators defined in the previous section, a contact-instanton defined in \cite{Kallen:2012aa} is a connection $A$ whose curvature $F_A$ satisfies
	\begin{equation}
		\pi _H^ + {F_A} = 0,\;\;{\pi _V}{F_A} = 0,
		\label{instanton}
	\end{equation}
or equivalently
	\begin{equation}
		{\pi ^ + }{F_A} = 0.
	\end{equation}
In \cite{Kallen:2012aa}, equations (\ref{instanton}) are discussed in the context of $(\kappa, R, g)$ being a K-contact structure. However, as also noted in \cite{Kallen:2012aa}, to define (\ref{instanton}), it it not necessary for $(\kappa, R, g)$ to be a contact structure. In the least constrained scenario, (\ref{instanton}) can be defined as long as $(\kappa, R, g)$ satisfies
	\begin{equation}
		\kappa(R) = 1, \;\;\;\;g(R, \cdot) = \kappa(\cdot).
	\end{equation}
	
	However, we want to relate such an object to Yang-Mills equation. as in 4-dimension self-dual instanton implies Yang-Mills equation. We now show that a more suitable geometry to consider is $(\kappa, R, g)$ being contact. Suppose $A$ satisfies (\ref{instanton}).  First note that
	\begin{equation}
		F_H^ -  = F \Rightarrow d_A F_H^- = 0.
	\end{equation}
Then when $\kappa$ is contact, we have
	\begin{equation}
		{d_A}*F =  - {d_A}\left( {\kappa  \wedge F_H^ - } \right) =  - d\kappa  \wedge F_H^ -  = 0,
	\end{equation}
where we have used the fact that $d\kappa \in \Omega_H^+ (M)$ (will be explained in appendix [\ref{Appendix-B}], and (\ref{vanishing-product}). Note that if $\kappa$ is not contact, the Yang-Mills equation is not guaranteed.

	For reasons that will be clear in a moment, let us introduce a slightly more generalized version of the above notion. Let $\phi$ be a section of $\Omega^0(M, \mathfrak{g})$, and as before $A$ the connection. We define contact instanton as a pair $(A, \phi) \in \mathcal{A} \times \Omega^0(M, \mathfrak{g})$:
	\begin{equation}
		\pi _V {F_A} = 0,\;\;\pi _H^ + {F_A} = \lambda \phi d\kappa, \;\;d_A \phi  = 0,\;\;\lambda \in \mathbb{R}.
		\label{contact-instanton}
	\end{equation}
We claim that this version of contact-instanton contains more interesting information about contact geometry, and is also related to generalized $\mathcal{N} =1$ supersymmetry of vector multiplet \cite{Pan:2013aa} and geometries where a globally defined Killing spinor is absent \cite{Harland:2011aa}. We will come back to this in a later section.
	
	When $\lambda = 0$, our notion of contact-instanton is the same as that in \cite{Kallen:2012aa}, except that $d_A \phi = 0$ is not part of the terminology there. However, for $\lambda \ne 0$, the contact instanton can only be defined with $(\kappa, R, g)$ satisfying an additional property
	\begin{equation}
		d\kappa  \in \Omega _H^ + \left( M \right),
	\end{equation}
namely, it must be horizontal and self-dual. This condition implies that at $p \in M$ where $d\kappa \ne 0$,
	\begin{equation}
		\kappa  \wedge d\kappa  \wedge d\kappa  =  d\kappa  \wedge *d\kappa  \ne 0,
	\end{equation}
and therefore $\kappa$ defines a contact structure (which will be introduced in section 3) on the region where $d\kappa \ne 0$. This is the first hint that  that contact-instanton with $\lambda \ne 0$ is closer related to contact geometry than its $\lambda = 0$ cousin.  However, we are still left with the cases where $d\kappa = 0$ at some points. This will be partially resolved when we introduce the cohomological field theory.

	The contact-instanton with $\lambda \ne 0$ also implies Yang-Mills equation when $A$ is irreducible. It is straight forward to show
	\begin{equation}
		{d_A}{F_A} = {d_A}F_H^ -  = 0,
	\end{equation}
and therefore
	\begin{equation}
		{d_A}*F = \lambda \phi d\kappa  \wedge d\kappa, 
	\end{equation}
where ${d_A}\phi = 0$ and (\ref{vanishing-product}) are used. When $\lambda$ or $\phi$ vanishes the Yang-Mills equation is recovered, and in particular, irreducible\footnote{A connection $A$ of associated vector bundle ${\rm ad} P_G$ is irreducible if the holonomy group $H_A = G$, and othersie if $H_A  < G$, $A$ is called a reducible connection. Let $\phi$ be a non-zero section such that $d_A \phi = 0$, namely $\phi$ corresponds to infinitesimal gauge transformation that leaves $A$ fixed, and therefore the Abelian subgroup of gauge transformations $g_t \equiv \exp(t\phi)$ also preserves $A$: ${g_t} \cdot A = A$. Let $\gamma_p$ be a loop based at $p \in M$. The holonomy $hol_A(\gamma)$ of $A$ along  transformed under gauge transformation as $ho{l_A}\left( \gamma  \right) = ho{l_{{g_t} \cdot A}}\left( \gamma _p \right) = {g_t}\left( p \right)ho{l_A}\left( \gamma_p  \right)g_t^{ - 1}\left( p \right)$. When $\gamma_p$ takes all the possible paths based at $p$, one sees that the Abelian subgroup commutes with $H_A$. Since we focus on semi-simple Lie group $G$, $H_A$ is forbidden to be the entire $G$, and therefore the existence of $\phi$ implies $A$ to be s reducible connection.
} contact-instanton with any $\lambda$ satisfies Yang-Mills equation.
	
	Another interesting property of contact-instanton is as follows. 
	
	Suppose now $A$ is reducible then there is non-zero solution to equation $d_A \phi = 0$. One can consider a set of 2-forms ${\text{tr}}\left( {\phi ^k F_H^ - } \right)$, and we have
	\begin{equation}
		d{\text{tr}}\left( {\phi ^k F_H^ - } \right) = {\text{tr}}\left( {{d_A}(\phi ^k)  \wedge F_H^ - } \right) + {\text{tr}}\left( {\phi ^k  {d_A}F_H^ - } \right) = 0,
	\end{equation}
and moreover,
	\begin{equation}
		{d^*}{\text{tr}}\left( {\phi^k  F_H^ - } \right) =  - *{\text{tr}}\left( {\phi^k  {d_A}\left( {\kappa  \wedge F_H^ - } \right)} \right)\; =  - *{\text{tr}}\left( {\phi^k  \left( {d\kappa  \wedge F_H^ - } \right)} \right) = 0,
	\end{equation}
where we have used ${d_A}\phi  = {d_A}F_H^ -  = 0$ as well as (\ref{vanishing-product}). This concludes that ${\text{tr}}\left( {\phi^k  F_H^ - } \right)$ defines a set of harmonic 2-form on $M$, which are all horizontal anti-self-dual.

	\section{Summary of contact geometry}
	
	In this section we will summarize relevant aspects of contact geometry that will be used in later sections. Interested readers may refer to appendix \ref{Appendix-B} for more details.
	
	\vspace{10pt}	
	\underline{Contact structure and Reeb vector field}
	
	Let $M$ be a $2n+1$-dimensional compact smooth manifold. The Euler number $\chi(M) = 0$ implies that $M$ admits nowhere-vanishing smooth vector fields or 1-forms. 
	
	Let $\kappa$ be a nowhere-vanishing 1-form. Then it defines a horizontal sub-bundle $T_H M$ of $TM$ by
	\begin{equation}
		{T_H}M \equiv \left\{ {\left( {p,X} \right) \in TM|{{\left. \kappa  \right|}_p}\left( X \right) = 0} \right\}.
	\end{equation}
	
	$\kappa$ defines a contact structure, or contact distribution $E = T_HM$ if
	\begin{equation}
		\kappa \wedge (d\kappa)^n \ne 0,
		\label{contact-condition}
	\end{equation}
everywhere on $M$. $\kappa$ itself is called a contact 1-form, and a manifold admitting a contact structure is called a contact manifold. 

	Once a contact 1-form is given, there is unique vector field $R$ such that
	\begin{equation}
		\kappa_m R^m = 1, \;\;\;\; R^m (d\kappa)_{mn} = 0.
	\end{equation}
and we call it the Reeb vector field associated to contact 1-form $\kappa$. The Reeb vector field on a compact contact manifold generates 1-parameter family of diffeomorphisms (an effective smooth $\mathbb{R}$-action on $M$), which is usually called the Reeb flow $\varphi_R(t)$, or the contact flow. The flow moves points along the integral curves of the Reeb vector field. It follows from the definition that the flow preserves $\kappa$: $\varphi_R^* \kappa = \kappa, \forall t$, or equivalently
	\begin{equation}
		\mathcal{L}_R \kappa = 0.
	\end{equation}

	The integral curves of Reeb vector fields have various types of behaviors. The regular type is that all the curves are closed and the Reeb flow generates free $U(1)$-action on $M$, rendering $M$ a principal $U(1)$-bundle over some symplectic $2n$-dimensional manifold. A quasi-regular type is that the curves are all closed but the flow only generates locally-free $U(1)$-action. The irregular type is that not all curves are closed, and they may have uncontrollable behaviors, which is the generic case.

	Note that one can rescale $\kappa \to e^f \kappa$ wile preserving (\ref{contact-condition}), and $E$ is invariant under such rescaling. Effective deformations of a contact structure $E$ come from $\delta \kappa \in \Gamma (T_HM^*)$. There is an important theorem by Gray, stating that if $\kappa_t$ is a family of contact structures, then they are all equivalent in the sense that there exists diffeomorphisms $\varphi_t : M\to M$ taking
	\begin{equation}
		\varphi _t^*{\kappa _0} = {e^{{h_t}}}{\kappa _t}.
	\end{equation}
for some family of real functions $h_t$, and therefore taking the corresponding contact distributions $\varphi _t^*{E_0} = {E_t}$.

	\underline{Contact metric structure}
	
	Given a contact 1-form $\kappa$, one can define a set of quantities $(\kappa, R, g, \Phi)$ where $g$ is a metric and $\Phi$ is a $(1,1)$-type tensor, such that
	\begin{equation}
		{g_{mn}}{R^n} = {\kappa _m},\;\;\;\;2 {g_{mk}}{\Phi ^k}_n = {\left( {d\kappa } \right)_{mn}} = {{\dot \nabla }_m}{\kappa _n} - {{\dot \nabla }_n}{\kappa _m}.
	\end{equation}
where $\dot \nabla $ denotes the Levi-civita connection of $g$. We call such set of quantities a contact metric structure.
	
	There are a few useful algebraic and differential relations between quantities. First we have
	\begin{equation}
		{\Phi ^n}_m{R^m} = {\kappa _n}{\Phi ^n}_m = 0,
	\end{equation}
	\begin{equation}
		\frac{{{{\left( { - 1} \right)}^n}}}{{{2^n}n!}}\kappa  \wedge {\left( {d\kappa } \right)^n} = {\Omega _g},
	\end{equation}
where $\Omega_g$ is the volume form associated to metric $g$. From this one can show that $d\kappa$ satisfies
	\begin{equation}
		{\iota _R}*d\kappa  = d\kappa.
	\end{equation}
	
	Moreover, we have
	\begin{equation}
		{R^n}{\nabla _m}{\kappa _n} = {\kappa _n}{\nabla _m}{R^n} = {R^m}{\nabla _m}{R^n} = 0,
	\end{equation}
which implies $R$ is geodesic.	

	There are useful relations between $R$ and $\Phi$. First we have
	\begin{equation}
		R^m \dot\nabla_m {\Phi^n}_k = 0.
		\label{Blair_1}
	\end{equation}
Also,
	\begin{equation}
		{\dot \nabla _m}{R^n} =  - {\Phi ^n}_m - \frac{1}{2}{\left( {\Phi ^\circ {\mathcal{L}_R}\Phi } \right)^n}_m.
		\label{Blair_2}
	\end{equation}
	
	\underline{K-contact structure}
	
	It is called a K-contact structure, if a contact metric structure satisfies an additional condition
	\begin{equation}
		\mathcal{L}_R g = 0.
	\end{equation}
Note that this is equivalent to
	\begin{equation}
		{\mathcal{L}_R}\Phi  = 0.
	\end{equation}
and consequently,
	\begin{equation}
		{\dot \nabla _m}{R^n} =  - {\Phi ^n}_m.
	\end{equation}

	Well-known examples of contact structures and K-contact structures are discussed in the appendix \ref{Appendix-B}. In particular, any quasi-regular contact structure always admits a contact metric structure which is K-contact \cite{Boyer:2012aa}.
	
	\section{Cohomological theory for contact-instanton}
	
	\subsection{The cohomological theory}
	
	Following the formalism of cohomological theory, namely the paradigm of ``fields, equation, symmetry", we consider the following set of fields: 
	\begin{itemize}
		\item connection 1-form $A$, with gauge symmetry, as usual;
		\item a even real scalar $\phi \in \Omega^0(M, \mathfrak{g})$;
		\item an odd 1-form $\psi_m \in \Omega^(M, \mathfrak{g})$;
		\item an even differential form $H \in \Omega^2_V(M, \mathfrak{g})\oplus\Omega^+_H(M, \mathfrak{g})$;
		\item an odd differential form $\chi \in \Omega^2_V(M, \mathfrak{g})\oplus\Omega^+_H(M, \mathfrak{g})$;
		\item an even real scalar $\bar \phi \in \Omega^0(M, \mathfrak{g})$;
		\item an odd real scalar $\eta \in \Omega^0(M, \mathfrak{g})$.
	\end{itemize}

	The BRST transformation $Q$ on the set of fields is defined as
	\begin{equation}
		\left\{ \begin{gathered}
		  QA = i\psi  \hfill \\
		  Q\phi  = 0 \hfill \\
		  Q\psi  = {d_A}\phi  \hfill \\
		  Q\chi_H^+  = H_H^+ - 2\lambda \phi d\kappa,\;\;Q\chi_V = H_V  \hfill \\
		  QH_H^+ = - \left[ {\phi ,\chi_H^+ } \right],\;\;QH_V = - [\phi, \chi_V] \hfill \\ 
		\end{gathered}  \right.,\;\;\;\;
		\left\{ \begin{gathered}
		  Q\bar \phi  = \eta  \hfill \\
		  Q\eta  = - \left[ {\phi ,\bar \phi } \right] \hfill \\ 
		\end{gathered}  \right..
		\label{Q-transformation}
	\end{equation}
which satisfies
	\begin{equation}
		{Q^2} = i{\mathcal{G}_\phi }.
	\end{equation}
where $\mathcal{G}_\Lambda$ denotes the gauge transformation with parameter $\phi$:
	\begin{equation}
		i \mathcal{G}_\Lambda A = i d_A \phi,\;\; i \mathcal{G}_\phi \Phi= -[\phi, \Phi],
	\end{equation}
for $\Phi$ any adjoint-valued fields.

	The Lagrangian we will consider is defined as follows:
	\begin{equation}
		\begin{gathered}
  		{\mathcal{L}}_\lambda = \frac{1}{{{e^2}}}{\text{tr}}\left[ {\frac{1}{2}F_H^ +  \wedge *F_H^ + +\frac{1}{4}{F_V} \wedge *{F_V} - \lambda \phi d\kappa  \wedge *F_H^ +  + {\lambda ^2}{\phi ^2} + {d_A}\bar \phi  \wedge *{d_A}\phi } \right. \hfill \\
		  \left. {\;\;\;\;\;\;\;\;\;\;\;\;\;\;\;\;\;\; - i \pi^+\chi  \wedge *{d_A}\psi  + i{d_A}\eta  \wedge *\psi  + \frac{1}{2} \pi^+\chi  \wedge *\left[ {\phi ,\chi } \right] + \psi  \wedge *\left[ {\bar \phi ,\psi } \right]} \right] \hfill ,\\
		\end{gathered} 
		\label{Lagrangian}
	\end{equation}
where $F_H^+$ and $F_V$ denotes the anti-self-dual and vertical part of curvature $F_A$ respectively. Note that the first two terms combines into
	\begin{equation}
		{F \wedge *F - \kappa  \wedge F \wedge F}.
	\end{equation}
which is the 5d analog of 4d Yang-Mills action and $\theta$ term.

	By direct computation, it is easy to verify that if we restrict ourselfes to conditions
	\begin{equation}
		\left\{ \begin{gathered}
		  \kappa \left( R \right) = 1,\;\;g\left( {R, \cdot } \right) = \kappa \left(  \cdot  \right)\;\;\;\;\;\;\;\;\;\;\;\;\;\;\;\;\;\;\;\;\;\;\;\;\;\;\;\;\;\;\;\;\;\;\;\;\;\;\;\;\;\;\;\;\;\;\;\;\;\;\;\;\;\;\;\;\;\;\;\;\;\;\;\;\;\;{\rm when\;}\lambda  = 0 \hfill \\[0.5em]
		  \kappa \left( R \right) = 1,\;\;g\left( {R, \cdot } \right) = \kappa \left(  \cdot  \right),\;\;d\kappa  \in \Omega _H^ + \left( M \right),\;\;{\left( {d\kappa } \right)_{mn}}{\left( {d\kappa } \right)^{mn}} = 4 \ne 0\;\;\;\;{\rm when\;}\lambda  \ne 0 \hfill \\ 
		\end{gathered}  \right.
		\label{conditions}
	\end{equation}
one has the partition function
	\begin{equation}
		Z_\lambda (\kappa, R, g; e) \equiv \int {D\left[ {A,\psi ,\phi ,\chi ,\bar \phi ,\eta } \right]{e^{ - {\mathcal{L}_\lambda }}}}  = \int {D\left[ {A,\psi ,\phi ,\chi ,\bar \phi ,\eta } \right]DH{e^{ - \left\{ {Q,V} \right\}}}}.
	\end{equation}
with
	\begin{equation}
		V = \frac{1}{{e^2 }}{\text{tr}}\left( {\frac{1}{2}\pi ^ -  \chi  \wedge *\left( {2F - H} \right) + d_A \bar \phi  \wedge *\psi } \right).
	\end{equation}
namely the Lagrangian (\ref{Lagrangian}) is equivalent to a $Q$-exact form. Let us mention that the second line of (\ref{conditions}) coincides with properties of a contact structure, which will be discussed in appendix [\ref{Appendix-B}], and sometime we will refer to it as ``contact conditions".

	It is straight forward to interpret terms in the Lagrangian. For obvious reasons the partition function is independent of finite non-zero $e$. The kinetic terms of $F_A$ and $\phi, \bar \phi$ force the integration over $DAD\phi$ to localize onto the contact instanton configuration (\ref{contact-instanton}), if one takes the integration contour $\bar \phi = + \phi$ and the weak coupling limit $e \to 0$. Integrating out $\chi$ enforces the deformation condition $\pi _H^ + {d_A}\psi  = 0$ and $\iota_R d_A\psi = 0$. Integrating out $\eta$ provides a gauge-fixing condition on $\psi$, namely
	\begin{equation}
		d_A^*\psi  = 0.
	\end{equation}
However, there are zero modes of $\chi$ satisfying
	\begin{equation}
		\pi _H^ - d_A^*\chi _H^ +  = 0,\;\;{\pi _V}d_A^*{\chi _V} = 0,
	\end{equation}
which accounts for the the second cohomology in the deformation complex (discussed later), while the zero modes
	\begin{equation}
		d_A^*{d_A}\bar \phi  = 0,
	\end{equation}
accounts for the zeroth cohomology of the complex, which signifies the reducibility of the connection $A$.
	
		One can add more $Q$-closed operators to the Lagrangian, namely
	\begin{equation}
		{I_k} \equiv \int_M {{\text{tr}}{\phi ^k}d\kappa  \wedge *d\kappa } ,
	\end{equation}
which serve as the classical contribution to the partition function.

	\underline{Remarks}
	
	Let us make a few remarks. We mentioned in previous section that $\lambda \ne 0$ may lead to more interesting information about contact geometry than $\lambda = 0$ case. This can be argued in the following way, concerning the property of $\delta \mathcal{L}_{\lambda}$ under deformation (or ``homotopy", in mathematical language) of background geometry $(\kappa, R, g)$.
	
	If $\lambda = 0$, the $Q$-transformation and the Lagrangian do not impose any additional condition on $d\kappa$, except for requiring $\kappa$ to unit norm: the transformation $Q$ can be defined ,and the action $Q$-exact for arbitrary unit-normed $\kappa$. Of course, it is completely fine to impose the condition $d\kappa \in \Omega_H^+(M)$ by hand, namely one can declare $\kappa$ to be a contact 1-form and  study (\ref{Lagrangian}). However, even if one starts with $\kappa$ being contact, the deformation $\delta \mathcal{L}_{\lambda = 0}$ under homotopy of $(\kappa, R, g)$ will be $Q$-exact even if $\kappa$ is deformed to be non-contact, as long as its norm remains $1$. In this sense, the theory provides homotopy invariants of hyperplane fields.
	
	On the contrary, when $\lambda \ne 0$, $d\kappa$ is already required to be self-dual by the particular transformation (\ref{Q-transformation}). Moreover, if one starts with contact background $(\kappa, R, g)$ satisfying (\ref{conditions}), $d\kappa  \cdot d\kappa  = 4$ must remain true along deformation of $(\kappa, R, g)$ if one wants $\delta \mathcal{L}_{\lambda \ne 0} = \{Q, ...\}$\footnote{Of course, if one starts with $d\kappa \cdot d\kappa = f$ with $f$ an arbitrary function, $\mathcal{L}_\lambda$ is still well-defined, and the $\delta \mathcal{L}_\lambda$ will be $Q$-exact if $d\kappa \cdot d\kappa = f$ keeps fixed under homotopy. However we focus on the case when $f = 4$ which is natural in contact geometry.}. In this sense, the theory with $\lambda \ne 0$ is sensitive enough to provide invariants of contact structures.
	
	Another reason of extending the definition is discussed in \cite{Harland:2011aa} to deal with the case where globally defined spinors is absent. The solution proposed in the example of K\"{a}hler manifold $M$, whose structure group is $U(\dim_\mathbb{C} M)$, bu including in the definition, known as Hermitian-Yang-Mills equation,
	\begin{equation}
		{F^{\mathfrak{u}\left( 1 \right)}} = \lambda \omega  \otimes J,
	\end{equation}
where $J$ is constant section of $End(E)$. This matches with our extended part, with $\omega$ matched with $d\kappa$ and $J$ with $\phi$. We do not however require $\phi$ to be central in this note.
	
	In the following sections, we will focus on $\lambda \ne 0$.

	\subsection{BRST Observables}
	
	The observables of this theory are obtained through descent equation, the same as those in Donaldson-Witten theory. One starts from $Q$-invariant observable $\mathcal{O}_k^{(0)} \equiv {\rm tr}\phi^k$,\footnote{the possible values of $k$ is determined by the gauge group $G$.} follows the descent equation
	\begin{equation}
		d\mathcal{O}_k^{\left( n \right)}  = \left\{ {Q,\mathcal{O}_k^{\left( {n + 1} \right)} } \right\},
	\end{equation}
and stops as one reaches $O_k^{(k)} = {\rm tr}F^k$. In this way, one can construct integrated operators
	\begin{equation}
		I_k^{\left( n \right)} (\Sigma^n)  = \int_{\Sigma ^n } {\mathcal{O}_k^{\left( n \right)} } ,
	\end{equation}
where $\Sigma^n$ denotes any $n$-cycle of $M$. The integrated observables $I_k^{(n)}$ is then $Q$-invariant but not $Q$-exact. As we will see, the expectation values of $I_k^{(n)}$ are invariant under homotopy of the underlying contact structure $(\kappa, R, g)$.

	For instance, when $k = 2$, we have
	\begin{equation}
		\begin{gathered}
		\mathcal{O}_2^{\left( 0 \right)}  = \frac{1}
		{{8\pi ^2 }}{\text{tr}}\phi ^2 ,\;\;\mathcal{O}_2^{\left( 1 \right)}  = \frac{1}
		{{4\pi ^2 }}{\rm tr}\phi \psi ,\;\;\mathcal{O}_2^{\left( 2 \right)}  =  - \frac{i}
		{{4\pi ^2 }}{\rm tr}\left( {\phi F + \frac{i}
		{2}\psi  \wedge \psi } \right) \hfill \\
  		\mathcal{O}_2^{\left( 3 \right)}  =  - \frac{i}
		{{4\pi ^2 }}{\rm tr}\left( {\psi  \wedge F} \right),\;\;\mathcal{O}_2^{\left( 4 \right)}  =  - \frac{1}
		{{8\pi ^2 }}{\text{tr}}\left( {F \wedge F} \right) \hfill. \\ 
		\end{gathered} 
	\end{equation}

	Notice that the operator $Q$ can be viewed as the differential operator on $\mathcal{M}$, and $A$ as the coordintates of $\mathcal{M}$. Therefore, the $Q$-complex indicates a way to assign degree of forms on $\mathcal{M}$ to the fields $({\bar \phi}, \chi, \eta, A, \psi, \phi)$ as $(-2, -1, -1, 0, +1, +2)$. Using this, one can interprete the observables $\mathcal{O}_k^{(n)}$ as differential forms defined on $\mathcal{M}$. Let us also mention that, as $\lambda \to 0$, the Lagrangian (\ref{Lagrangian}) has an additional $U$-symmetry under which the charges of each fields equal their degree of forms on $\mathcal{M}$. In that case, the anomaly cancellation condition automatically selects the top degree part of the observable to integrate over $\mathcal{M}$.\footnote{When $\lambda \ne 0$, (\ref{Lagrangian}) also has the same $U$-symmetry if we assign charge $-2$ to the constant $\lambda$. } With this in mind, only the observables that have total degrees equal to $\dim \mathcal{M}$ give non-zero expatiation values. On geometry satisfying certain restrictions, $\dim \mathcal{M}$ can be read off from the $\phi_i$ independent term of the integral
	\begin{equation}
		{\text{in}}{{\text{d}}_{{T^k}}}D_V^ + \left( {{e^{i{\varphi _1}}},...,{e^{i{\varphi _k}}}} \right) = {\left( {\frac{1}{{2\pi i}}} \right)^2}\int_M {Td\left( {T_H^{0,1}M} \right)\left( {{\varphi _i}} \right) \wedge \mathcal{J}\left( \kappa  \right)\left( {{\varphi _i}} \right) \wedge Ch(V,A)\left( {{\varphi _i}} \right)} .
	\end{equation}

	\subsection{Homotopy invariance}
	
	Now let us consider what happens if the data $(\kappa, R, g)$ is deformed. As expected, the resulting deformation of (\ref{Lagrangian}) is again $Q$-exact.
	
	Suppose we start with $(\kappa, R, g)$ satisfying
	\begin{equation}
		\kappa (R)  = 1,\;\;\;\;g\left( {R, \cdot } \right) = \kappa \left(  \cdot  \right).
	\end{equation}
	
	As we deform 1-form $\kappa$ by $\delta\kappa$, $g$ and $R$ may also  need to deform by $\delta g$ and $\delta R$ to maintain the conditions above. The deformations must satisfy
	\begin{equation}
		{R^m}\delta {\kappa _m} + \delta {R^m}{\kappa _m} = 0,\;\;\;\;\delta {g_{mn}}{R^n} + {g_{mn}}\delta {R^n} = \delta {\kappa _m}.
	\end{equation}

	However, we know that deformation of the form $\delta \kappa \propto \kappa $ does not change the hyperplane field, which is the more essential geometric object. Hence, let us consider effective change of 1-form $\kappa$, namely
	\begin{equation}
		{\iota _R}\delta \kappa  = 0,
	\end{equation}
and therefore
	\begin{equation}
		\delta {g_{mn}}{R^m}{R^n} = 0.
	\end{equation}
	
	The deformation also changes the notion of self-dual 2-forms and vertical 2-forms. Let $\omega$ be an arbitrary anti-self-dual 2-form, then to maintain self-duality it must deform as
	\begin{equation}
		\begin{gathered}
		  \delta {\omega _{mn}} =  - \frac{{{g_{pq}}\delta {g^{pq}}}}{4}{\omega _{mn}} \hfill \\
		  \;\;\;\;\;\;\;\;\;\;\;\;\;  \frac{{\sqrt g }}{4}\left( {\delta {g^{kl}}{\kappa _l}{\epsilon _{kmnpq}}{\omega ^{pq}} + {g^{kl}}\delta {\kappa _l}{\epsilon _{kmnpq}}{\omega ^{pq}} + 2\delta {g^{pp'}}{g^{qq'}}{R^k}{\epsilon _{kmnp'q'}}{\omega _{pq}}} \right) \hfill \\
		  \;\;\;\;\;\;\;\;\; \equiv {\left( {{L_{\delta \kappa ,\delta g,\delta R}}\omega } \right)_{mn}}, \hfill \\ 
		\end{gathered} 
	\end{equation}
where we have defined a complicated linear operator $L_{\delta \kappa, \delta R, \delta g}$ which acts on self-dual 2-forms by contracting indices. Note that if field $\omega$ is self-dual, $\{Q, \omega \}$ is also self-dual, then we have
	\begin{equation}
		\delta \left\{ {Q,\omega } \right\} = L\left\{ {Q,\omega } \right\} = \left\{ {Q,L\omega } \right\} = \left\{ {Q,\delta \omega } \right\}.
	\end{equation}
	
	Similarly, let $\mu$ be arbitrary vertical 2-form, then it deforms as
	\begin{equation}
		\delta \mu  =  - \left( {{\iota _{\delta R}}\mu } \right)\kappa 
	\end{equation}.
Note that we also have for vertical forms
	\begin{equation}
		\delta \left\{ {Q,\mu } \right\} = \left\{ {Q,\delta \mu } \right\}.
	\end{equation}
	
	Combining with the fact that (\ref{Lagrangian}) is $Q$-exact, one sees that under the deformation, (\ref{Lagrangian}) changes by $Q$-exact amount, since relevant terms in (\ref{Lagrangian}) are of the form
	\begin{equation}
		\left\{ {Q,{\omega _1} \wedge *{\omega _2}} \right\}\;\;\;\;{\text{or}}\;\;\;\;\left\{ {Q,{\mu _1} \wedge *{\mu _2}} \right\},
	\end{equation}
and under deformation
	\begin{equation}
		\delta \left\{ {Q,{\omega _1} \wedge *{\omega _2}} \right\} = \left\{ {Q,\delta {\omega _1} \wedge *{\omega _2} + {\omega _1} \wedge *\delta {\omega _2} + \delta {g^{mm'}}{g^{nn'}}{{\left( {{\omega _1}} \right)}_{mn}}{{\left( {{\omega _2}} \right)}_{m'n'}}} \right\},
	\end{equation}
and similar for vertical $\mu$'s. Therefore, the partition function $Z$ is actually invariant under all smooth deformation of the data $(\kappa, R, g)$ which keeps the following relation
	\begin{equation}
		\kappa(R)  = 1,\;\;g\left( {R, \cdot } \right) = \kappa \left(  \cdot  \right).
	\end{equation}
In this sense, the partition function is an invariant of hyperplane field defined by $\kappa$.

	Recall that in additional to the horizontal condition, $\kappa$ has to satisfy a few more contact constraints in (\ref{conditions}). However these dose not alter the final statement, that the deformation of the theory is again $Q$-exact. If one starts with the geometry satisfying (\ref{conditions}), any deformation violating  (\ref{conditions}) with result in non-$Q$-exactness of the deformation of Lagrangian (\ref{Lagrangian}), and therefore in general the partition function will change under deformation. In this sense, the partition function $Z(\kappa, R, g; e)$ is only guaranteed to be invariant under deformations satisfying constraints (\ref{conditions}), which coincides happily with the deformation of contact structure, as we will review in the next section.
	
	Recall that we have a set of $Q$-invariant observables determined by the gauge group $G$. Therefore, we have for any set of cycles $\Sigma_i^{n_i}$ of $M$, we have invariants
	\begin{equation}
		{F_{O_{{k_i}}^{{n_i}},\Sigma _i^{{n_i}}}}\left( {\kappa ,R,g} \right) \equiv \left\langle {\prod\limits_i {I_{{k_i}}^{\left( {{n_i}} \right)}\left( {\Sigma _i^{{n_i}}} \right)} } \right\rangle _{(\kappa, R, g)}  = \left\langle {\prod\limits_i {\int_{\Sigma _i^{{n_i}}} {\mathcal{O}_{{k_i}}^{\left( {{n_i}} \right)}} } } \right\rangle _{(\kappa, R, g)},
	\end{equation}
that are invariant under homotopy of contact structure $(\kappa, R, g)$. In particular, when the moduli space of contact-instanton are discrete points, the partition function reduces to a counting quantitiy
	\begin{equation}
		Z = \sum\limits_i {{{\left( { - 1} \right)}^{{n_i}}}} ,
	\end{equation}
where $i$ labels the contact-instanton solutions.

	\section{The contact-instanton moduli $\mathcal{M}_\lambda$}
	
	\subsection{The deformation complex and vanishing theorem}
	
	In this section we will study the dimension of instanton moduli space $\mathcal{M}_\lambda$. For general contact metric structures, we do not have straight forward way to compute $\dim \mathcal{M}$. However, on some specially simple sectors, the computation become straightforward. We will mainly consider irreducible connections with a specific gauge choice. and focus on K-contact structures, to make computation possible.

	\vspace{20pt}
	The deformation of (\ref{contact-instanton}) reads
	\begin{equation}
		{\pi _V}{d_A}\delta A = 0,\;\;\pi _H^ - {d_A}\delta A = \lambda \delta \phi d\kappa ,\;\;{d_A}\delta \phi  - i [\delta A, \phi] = 0.
		\label{original_deformation}
	\end{equation}
where the data $(\kappa, R, g)$ satisfy (\ref{conditions}) for $\lambda \ne 0$ with contact 1-form $\kappa$.
	
	The deformation complex so defined is not elliptic, and there is no rigorous way to study the index of corresponding operators for the most general case. Therefore, let us try to understand some simplest cases.

	 Let us consider the case when $A$ is an irreducible connection. Then $\phi = 0$ and one is left with the equation
	\begin{equation}
		{\pi ^ - }{d_A}\delta A = 0
		\label{original-instanton-deformation}.
	\end{equation}

	Note that the condition (\ref{contact-instanton}) requires $F$ to be horizontal, or equivalently,
	\begin{equation}
		{\mathcal{L}_R}A = {d_A}\left( {{\iota _R}A} \right).
	\end{equation}
This implies that an infinitesimal action of diffeomorphism generated by $R$ is equivalent to infinitesimal gauge transformation with parameter $\iota_R A$. In some sense, the condition allows to ``gauge away" $\mathcal{L}_R A$ to make it vanish\footnote{Note that there are global obstructions to such gauge.}.

	The dimension of irreducible contact-instanton can be viewed as the dimension of the first cohomology of the following complex
	\begin{equation}
		0 \xrightarrow{\iota} \Omega^0 (M, \mathfrak{g}) \xrightarrow{d_A} \Omega^1(M, \mathfrak{g})  \xrightarrow{ \pi^+ d_A} \Omega^2_V(M, \mathfrak{g}) \oplus \Omega _H^ + \left(M, {\mathfrak{g}} \right) \xrightarrow{\pi_H^-} 0.
	\end{equation}
	
	The complex is neither elliptic nor transversally elliptic, and therefore the dimensions of the cohomologies are in general not well-defined. However, we can look at some sectors where various quantities are better defined.
	
	Let us first assume that for an irreducible contact instanton $A$ we can make a gauge choice such that
	\begin{equation}
		\mathcal{L}_R A = 0.
	\end{equation}
By irreducibility this implies $\iota_R A = 0$. There are residual gauge freedom that needs to be modded out, namely $\tilde \phi \in \tilde \Omega^0 (M, \mathfrak{g})$ such that
	\begin{equation}
		\mathcal{L}_R \tilde \phi = 0.
	\end{equation}
We then have a reduced deformation complex:
	\begin{equation}
		0 \xrightarrow{\iota}\tilde \Omega^0 (M, \mathfrak{g}) \xrightarrow{d_A} \tilde\Omega^1(M, \mathfrak{g})  \xrightarrow{ \pi^+ d_A}  \tilde\Omega _H^ + \left(M, {\mathfrak{g}} \right) ,\xrightarrow{\pi_H^-} 0,
		\label{reduced-deformation-complex}
	\end{equation}
where $\tilde \Omega^*(M, \mathfrak{g})$ denotes basic differential forms that satisfy
	\begin{equation}
		{\iota _R}\tilde \omega  = {\mathcal{L}_R}\tilde \omega  = 0.
	\end{equation}
Note that if $\tilde \psi  \in {\tilde \Omega ^1}\left( {M,\mathfrak{g}} \right)$, then ${\pi ^ + }{d_A}\tilde \psi  \in \tilde \Omega _H^ + \left( {M,\mathfrak{g}} \right)$ without the vertical ingredients. Define the corresponding differential operator
	\begin{equation}
		{\tilde D_A} \equiv d_A^* + {\pi ^ + }{d_A}:{\tilde \Omega ^1}\left( {M,\mathfrak{g}} \right) \to {\tilde \Omega ^0}\left( {M,\mathfrak{g}} \right) \oplus \tilde \Omega _H^ + \left( {M,\mathfrak{g}} \right).
	\end{equation}

	The dimension of the moduli space of contact-instanton $A$ with the gauge choice above equals the dimension of the first cohomology of this reduced complex:
	\begin{equation}
		{T_A}{\mathcal{M}_{\lambda }} = \frac{{\ker {\pi ^ + }\circ {d_A}:{{\tilde \Omega }^1}\left( {M,\mathfrak{g}} \right) \to \tilde \Omega _H^ + \left( {M,\mathfrak{g}} \right)}}{{\operatorname{Im} {d_A}:{{\tilde \Omega }^0}\left( {M,\mathfrak{g}} \right) \to {{\tilde \Omega }^1}\left( {M,\mathfrak{g}} \right)}} \equiv \tilde H^1.
	\end{equation}
There is no direct way to obtain $\dim \tilde H^1$, and to compute $\dim \tilde H^1$, let us consider the following complex
	\begin{equation}
		0\xrightarrow{\iota }{\Omega ^0}(M,\mathfrak{g})\xrightarrow{{{d_A}}}\Omega _H^1(M,\mathfrak{g})\xrightarrow{{\pi _H^ +  \circ  {d_A}}}\Omega _H^ - \left( {M,\mathfrak{g}} \right)\xrightarrow{{\pi _H^ - }}0.
		\label{complex_3}
	\end{equation}

	Recall that there are 3 types of contact structure. corresponding to whether or not the integral curves of $R$ are closed. Now suppose the contact metric structure is actually a K-contact structure, namely a contact metric structure with an additional condition $\mathcal{L}_R g = 0$ satisfied. Then $R$ generates a 1-parameter subgroup of isometry group of $M$, and therefore its closure is a torus $T^k$ isometric action on $M$\cite{Meyer}\footnote{$k$ can take values from 1 to 3, on a 5-dimensional K-contact manifold. In general, on a $2n+1$-dimensional K-contact manifold, $k$ takes values between 1 and $n+1$, and the $n+1$ case is usually called ``completely integrable".}, namely, $\mathcal{L}_{T^k} g = 0$. Further more, since the flows of $R$ are dense on $T^k$, any $t \in T^k$ can be approximated by some flow of $R$, and the whole K-contact structure is invariant under the $T^k$ action (actually, any quantity that is invariant under $R$ will be invariant under the $T^k$-action). Moreover, any operator that commutes with $\mathcal{L}_R$, will commute with $T^k$-action:
	\begin{equation}
		{\pi _V}{\mathcal{L}_{T^k}} = {\mathcal{L}_{T^k}}{\pi _V},\;\;\left( {{\iota _{T^k}}*} \right){\mathcal{L}_{T^k}} = {\mathcal{L}_R{T^k}}\left( {{\iota _{T^k}}*} \right),\;\;{d_A}{\mathcal{L}_{T^k}} = {\mathcal{L}_{T^k}}{d_A}.
	\end{equation}
In the following, unless explicitly stated, we will focus on K-contact structures.
	
	One then concludes that the complex (\ref{complex_3}) above is actually transversally elliptic with respect to the $T^k$-action, and we have the corresponding transversally elliptic operator as
	\begin{equation}
		{D_A}:\Omega _H^1\left( {M,\mathfrak{g}} \right) \to {\Omega ^0}\left( {M,\mathfrak{g}} \right) \oplus \Omega _H^ + \left( {M,\mathfrak{g}} \right), \;\;{D_A} \equiv d_A^* + \pi _H^ + \circ {d_A},
	\end{equation}
which commutes with $\mathcal{L}_{T^k}$, and the $T^k$-equivariant index of $D_A$: 
	\begin{equation}
		{\text{in}}{{\text{d}}_{{T^k}}}{D_A}\left( g \right) = \sum\limits_{\rho  \in {\text{irrep}}} {\left[ {{{\left. {\chi \left( g \right)} \right|}_{{{\left( {\ker {D_A}} \right)}_\rho }}} - {{\left. {\chi \left( g \right)} \right|}_{{{\left( {{\text{coker}}{D_A}} \right)}_\rho }}}} \right]} ,\;\; g \in T^k.
	\end{equation}
where we decompose spaces ${\rm ker}D_A$ and ${\rm coker}D_A$ into irreducible representations of $T^k$:
	\begin{equation}
		\ker {D_A} = \mathop  \oplus \limits_\rho  {\left( {\ker {D_A}} \right)_\rho },\;\;\;\;{\text{coker}}{D_A} = \mathop  \oplus \limits_\rho  {\left( {{\text{coker}}{D_A}} \right)_\rho }.
	\end{equation}
	
	Note that the basic differential forms $\tilde \Omega^*(M, \mathfrak{g})$ are invariant under $T^k$ and therefore we can identify ${\rm ind}\tilde D_A$ with the coefficient of the term independent of $g$ in ${\rm ind}_{T^r} D_A (g)$. Hence, if we can compute ${\rm ind}_{T^k} D_A$ together with some vanishing theorem to ensure vanishing adjoint kernel
	\begin{equation}
		\ker \tilde D_A^* = \ker \left( {{d_A} + {{\left( {\pi _H^ + {d_A}} \right)}^*}} \right) = \ker {\left( {\pi _H^ + {d_A}} \right)^*} ={\rm ker}(d_A^* \pi^+_H) = 0,
	\end{equation}
where the irreducibility of $A$ has been used, then we can obtain
	\begin{equation}
		\dim {T_A}{\mathcal{M}_{\lambda}} = \dim {H^1} = {\text{ind}}{{\tilde D}_A}.
	\end{equation}

	Let us end this section by showing under what condition the vanishing theorem holds true. Let $\omega \in \tilde \Omega_H^+(M,\mathfrak{g})$ be a solution to equation
	\begin{equation}
		{\left( {\pi _H^ + {d_A}} \right)^*}\omega  = d_A^*\pi _H^ + \omega  = d_A^*\omega  = 0,
	\end{equation}
and as usual, we consider the equivalent Laplace equation,
	\begin{equation}
		\pi _ +  d_A d_A^* \omega  = 0.
	\end{equation}
We first compute the second-order differential operator:
	\begin{equation}
		d_A d_A^* \omega  = \iota _R *d_A^* d_A \omega  + d_A *\left( {d\kappa  \wedge \omega } \right) + {\cal L}_R *d_A \omega  - i\left[ {\iota _R A,*d_A \omega } \right].
	\end{equation}
With the aid of gauge choice $\iota_R A = 0$, the Killing condition $\mathcal{L}_R * = * \mathcal{L}_R$ and $(\iota_R *)^2 \alpha = \alpha - \alpha_V, \forall \alpha \in \Omega^2(M)$, we have
	\begin{equation}
		\iota _R *d_A d_A^* \omega  = d_A^* d_A \omega  - \left( {d_A^* d_A \omega } \right)_V  + \iota _R *d_A *\left( {d\kappa  \wedge \omega } \right).
	\end{equation}
Adding the term $\left( {d_A d_A^* \omega } \right)_H  = d_A d_A^* \omega  - \left( {d_A d_A^* \omega } \right)_V $ on both sides, and taking the inner product with $\omega$ itself, one obtaines
	\begin{equation}
		\left( {\omega ,\pi _H^ +  d_A d_A^* \omega } \right) = \left( {\omega ,d_A d_A^* \omega  + d_A^* d_A \omega } \right) + \left( {\omega ,\iota _R *d_A *\left( {d\kappa  \wedge \omega } \right)} \right).
	\end{equation}
The last term actually vanishes\footnote{A side remark: Let $\omega \in \Omega_H^+(M, \mathfrak{g}) \cap \ker d_A^*$, we have $*\left( {d\kappa  \wedge \omega } \right) = N\left( {d\kappa _{pq} \omega ^{pq} } \right)\kappa $ for some constant $N \ne 0$, and therefore
	\begin{equation}
		d_A *\left( {d\kappa  \wedge \omega } \right) = Nd_A \left( {d\kappa _{pq} \omega ^{pq} } \right) \wedge \kappa  + N\left( {d\kappa _{pq} \omega ^{pq} } \right)d\kappa .
	\end{equation}
Now that the inner product with $\omega$ itself vanishes, one obtains
	\begin{equation}
		\int {N\left( {d\kappa _{pq} \omega ^{pq} } \right)\omega  \wedge *d\kappa }  \propto \int {\left[ {\left( {d\kappa } \right)_{mn} \omega ^{mn} } \right]^2 }  = 0.
	\end{equation}
which implies $(d\kappa)_{mn} \omega^{mn} = 0$.
	}, since $\iota_R * $ is self-adjoint and $\iota_R * \omega = \omega$:
	\begin{equation}
		\left( {\omega ,d_A *\left( {d\kappa  \wedge \omega } \right)} \right) = \left( {d_A^* \omega ,*\left( {d\kappa  \wedge \omega } \right)} \right) = 0,
	\end{equation}
and finally we have an identify
	\begin{equation}
		\left( {\omega ,\pi _ +  d_A d_A^* \omega } \right) = \frac{1}{2}\left( {\omega ,d_A^* d_A \omega  + d_A d_A^* \omega } \right),
	\end{equation}
which closely resemble the 4-dimensional analog.

	We are left with the Laplacian operator $\Delta_A$. It is then straight forward to show that, when metric $g$ has scalar curvature $\mathcal{R}_g>0$, and the induced maps $Ric_g:\Omega _H^ + \left( {M,adP} \right) \to \Omega _H^ + \left( {M,adP} \right)$ and $W_g:\Omega _H^ + \left( {M,adP} \right) \to \Omega _H^ + \left( {M,adP} \right)$ defined as\footnote{$W_{mnkl}$ is the Weyl tensor of $g_{mn}$}
	\begin{equation}
		Ric_g \left( \omega  \right) \equiv \pi _H^ + \left( {{R_m}^k{\omega _{kn}}d{x^m} \wedge d{x^n}} \right),
	\end{equation}
	\begin{equation}
		W_g \left( \omega  \right) \equiv \pi _H^ + \left( {{W_{mnkl}}{\omega ^{kl}}d{x^m} \wedge d{x^n}} \right)
	\end{equation}
are positive semi-definite and non-positive respectively, the strict positivity
	\begin{equation}
		\left( {\omega ,{\Delta _A}\omega } \right) = 0 \Leftrightarrow \omega  = 0
	\end{equation}
can be achieved. Note that these conditions of metric are satisfied by any Sasaki-Einstein structure: on Sasaki-Einstein 5-manifolds,
	\begin{equation}
		\mathcal{L}_R g = 0,\;\;Ric_{mn}  = 4g_{mn} ,\;\;\mathcal{R} = 20,\;\;\iota _R *W\left( \omega  \right) =  - W\left( \omega  \right), \;\; \forall \omega \in \Omega^2(M).
	\end{equation}

	To summarize, we make use of $\mathcal{L}_R A = d_A (\iota_R A)$ to make a gauge choice such that $\mathcal{L}_R A = 0$, and we also focus on irreducible connections to ensure $\iota_R A = 0$. Then we can use an improved deformation complex to compute $\dim \mathcal{M}$. The Killing condition $\mathcal{L}_R g = 0$ does three things: it modifies the original $\mathbb{R}$-action\footnote{The smooth $\mathbb{R}$-action generated by the Reeb vector field $R$; since $M$ is compact and smooth, and $R$ is smooth, therefore $R$ is a complete vector field, and it generates a smooth map $\varphi_R: \mathbb{R} \times M \to M$.} to a compact group $T^k$-action on $M$, provides commutivity $D_A\mathcal{L}_R = D_A \mathcal{L}_R$, and gives rise to the vanishing theorem.
	
	\subsection{Spinor bundle and Dirac operator}
	
	To use known mathematical results, we first need to reinterprete the operator that we encountered. And it turns out that similar to 4-dimensional instanton study, $\pi_H^+ \circ d_A + d^*$ is related to some canonical Dirac-like operator defined on any contact metric manifold. In this section we will focus on the identification of the operator $\pi_H^+ \circ d_A + d^*$ with the twisted Dirac operator $\slashed{D}_{S_+^* \otimes {\rm ad}P_G}^-$, and interested readers may refer to Appendix [\ref{Appendix-B}][\ref{Appendix-C}], where we provide more pedagogical review on the construction of the spinor bundle as well as the Kohn-Dirac operator $\slashed{D}$.
	
	Let $S = S_+ \oplus S_-$ be the canonical spinor bundle over contact metric manifold $M$, where
	\begin{equation}S_+ = \wedge^{0,0}T_HM^* \oplus \wedge^{0,2} T_HM^*,\;\;\;\;S_- = \wedge^{0,1}T_HM^*.
	\end{equation}
	
	Let $\slashed{D}^- : \Gamma(S_-) \to \Gamma(S_+)$ be the Kohn-Dirac operator defined via generalized Tanaka-Webster connection $\nabla$ (we denote the Levi-civita connection by $\dot \nabla$). Let us consider first twisting the bundles with $S_+^*$, and consider twisted Kohn-Dirac operator 
	\begin{equation}
		\slashed{D}_{S_+^*}^- : \Omega_H^1(M)_\mathbb{C} \to \Omega^+_H(M)_\mathbb{C}\oplus \Omega^0(M)_\mathbb{C}.
	\end{equation}
Let $\psi \in \Omega_H^1(M_\mathbb{C})$, and then we have (\ref{Kohn-Dirac_2})
	\begin{equation}
		\slashed{D}_{S_ + ^* }^ -  \psi  = \sum\limits_a {E^a  \cdot \nabla _{E_a } \psi } = \sum\limits_a {\pi _H^ +  \left( {E^a  \wedge \nabla _{E_a } \psi } \right) - \iota \left( {E_a } \right)\nabla _{E_a } \psi }  ,\;\; \forall \psi \in \Omega_H^1(M)_\mathbb{C}.
		\label{Kohn-Dirac}
	\end{equation}
	
	Let us try to understand what this operator is. 
	
	The second term in (\ref{Kohn-Dirac}) is easy to reinterprete. It equals
	\begin{equation}
		- \iota \left( {E_a } \right)\nabla _{E_a } \psi  - \iota \left( R \right)\nabla _R \psi  + \iota \left( R \right)\nabla _R \psi  = d^* \psi  + R^n R^m \nabla _m \psi _n  = d^* \psi .
	\end{equation}
In the first equality, we use the completeness relation $\sum\limits_a {E_a^m E_n^a }  = \delta _n^m  - \kappa _n R^m $ and apply explicit expression of the contorsion ${C^k}_{mn}$ and find that the terms with $C$ vanish following from (\ref{Blair_1}) and (\ref{Blair_2}):
	\begin{equation}
		g^{mn} {C^k} _{mn} \psi _k  = \left( {\kappa ^m {\Phi ^k} _m  + \kappa ^m (\Phi  + \frac{1}{2}\Phi {\cal L}_R \Phi )^k \:_m } \right)\psi _k  = 0.
	\end{equation}
We then use
	\begin{equation}
		- \dot \nabla _m \psi ^m  = d^* \psi .
	\end{equation}
The second equality comes from the fact that $\nabla$ preserves $R$, see (\ref{GTW_1}).
	
	The first term in (\ref{Kohn-Dirac}) is also straightwoard. The result is
	\begin{equation}
		\sum\limits_a {E^a  \wedge \nabla _{E_a } \psi }  = d\psi  - \kappa  \wedge \iota _R d\psi  = \iota _R \left( {\kappa  \wedge d\psi } \right) = \pi _H d\psi.
	\end{equation}

	Combining the two terms, we see that the Kohn-Dirac operator (\ref{Kohn-Dirac_2}) actually equals to
	\begin{equation}
		\slashed{D}_{S_+^*} ^- \psi = \pi_H^- d\psi + d^* \psi, \;\;\;\;\forall \psi\in\Omega_H^1(M)_\mathbb{C}.
	\end{equation}
	
	This is almost the operator that we considered in the maintext. What remains is further twisting the bundle by vector bundle ${\rm ad}P_G$ with connection $A$, and we obtain the desired operator
	\begin{equation}
		\slashed{D}^-_{S_+^* \otimes {\rm ad}P_G} = \pi_H^+ \circ d_A + d_A^* = D_A.
	\end{equation}
Note that this identification is valid for any contact metric structure, without imposing $\mathcal{L}_R g = 0$.
	
	\subsection{Computing $\dim \mathcal{M}$ via equivariant index theorem}
	
	In the last two sections, we have shown that the dimension of contact-instanton moduli is related to the equivariant index of the operator $D_A = \pi_H^+ d_A + d_A^*$, which can be identified as the Kohn-Dirac operator $\slashed{D}^-_{S_+^* \otimes {\rm ad}P_G}$, on any contact metric manifold, and we will just write $D_A$ for simplicity.
	
	Although the identification of $D_A$ with a Kohn-Dirac operator is valid on general contact metric manifold, there is no natural transversal ellipticity. However, the situation is improved when we focus on K-contact manifold. There the induced isometric $T^k$-action preserves the contact  metric structure, and commutes with $D_A$. Moreover, the principal symbol of $D_A$ is invertible along orthogonal complement of $T^k$-action, and therefore is transversally elliptic in the sense of Atiyah\cite{Atiyah}.
	
	The equivariant index of transversally elliptic operator on contact manifold with elliptic $G$-action is computed in \cite{Fitzpatrick:2007aa}. We will review some relevant materials in this section, and apply them to our case.
	
	\vspace{20pt}
	\underline{Transversally elliptic operators on contact manifolds}
	
	Let $(M, \kappa, R)$ be a contact manifold, and $T_H M$ the contact distribution (horizontal tangent bundle). Let Lie group $G$ acts on $M$ smoothly, and preserves the contact structure. In particular,
	\begin{equation}
		{g^*}\kappa  \propto \kappa , \;\;\;\; \forall g\in G.
	\end{equation}
	
	The $G$-action is said to be elliptic if the orbits of $G$ in $M$ are nowhere tangent to the bundle $T_HM$. Denote $TG_p \subset T_p M$ be the tangent subspace tangent to the $G$-orbit at $p$, and define cotangent subspace $T_GM_p^* \subset T_pM^*$ as
	\begin{equation}
		T_GM_p^* \equiv \left\{ {\left( {p,\xi } \right) \in {T^*}M|\xi \left( T \right) = 0,\;\;\forall T \in T{G_p}} \right\},
	\end{equation}
namely space of 1-forms that gives zero on $TG_p$. Then ``the $G$-action is elliptic" is equivalent to ${\left. {{R_p}} \right|_{{T_G}M_p^*}} = 0, \forall p$. In particular, this implies ${T_G}M_p^* \subset {\left. {{T_H}{M^*}} \right|_p}$.
	
	1) For example, let $M = S^1$ with canonical contact structure $(\kappa = d\theta, R = \partial_\theta)$, and let $G = U(1)$ acts on $S^1$ in the obvious manner. Then $T_GM^* = 0$, since the subspace $TG_p = T_pM$ at each $p\in M$, and there is no more direction that is orthogonal to it. Therefore the $U(1)$ action is automatically elliptic, since condition ${\left. {{R_p}} \right|_{{T_G}M_p^*}} = 0$ does not say anything meaningful and the action obvious preserves $\kappa$.
	
	2) Another example would be $M = S^{2k+1}, k > 0$ with the contact structure given by Hopf-fibration over $\mathbb{C}P^k = S^{2k+1} / U(1)$. Let $G = U(1)$-action be the $U(1)$ in the Hopf-fibration, and therefore $R$ is tangent to the $U(1)$ orbit, $TU\left( 1 \right) = \mathbb{R} R $. The 1-forms orthogonal to $T_{U(1)}M_p^*$ are then horizontal 1-forms, ${T_{U\left( 1 \right)}}{M^*} = {T_H}{M^*}$. Therefore ${\left. R \right|_{{T_{U\left( 1 \right)}}{M^*}}} = 0$ and the $U(1)$-action is elliptic.
	
	Let $D$ be some differential operator on $M$ with $G$-action, and the principal symbol $\sigma_D(p, \xi)$, where $(x, \xi )\in TM^*$. Then $D$ is said to be transversally elliptic with respect to the $G$-action, if $\sigma_D (p, \xi)$ is invertible whenever $\xi \in T_G M^*_p$ for all $ p$.
	
	As an example, let us again consider first $M = S^1$ and $G = U(1)$ as above. Since $T_GM_p^* = 0$, there is no actual requirement for any differential operator $D$ on $S^1$ to be transversally elliptic, namely, even the highly degenerate zero operator $0$ is transversally elliptic.
	
	More generally, let $\slashed{D} : S \to S$ be the Kohn-Dirac operator defined on contact metric manifold $M$, and let $G$ acts on $M$ elliptically. Then the principal symbol $\sigma_\slashed{D} (p, \xi) = -i \xi \cdot $, where $\cdot$ denote the Clifford multiplication on $S$. Note that only the horizontal part of $\xi$ acts non-trivially on $S$, and therefore $\sigma_\slashed{D} (p, \xi) = - {\left| {{\xi _H}} \right|^2}$, which is non-zero for $\xi_H \ne 0$. This implies $\sigma_\slashed{D}(p, \xi)$ is invertible on $T_HM^*$, and therefore also invertible on the subspace $T_pM^*_G$. If further more the operator $\slashed{D}$ commutes with $G$-action, $\slashed{D}$ is transversally elliptic.

	\vspace{20pt}
	\underline{Equivariant differential forms}
	
	Let $\mathfrak{g}$ be the Lie algebra of $G$ which acts on contact manifold $M$ elliptically. Then each $T \in \mathfrak{g}$ induces a vector field $T_M$ on $M$ representing the infinitesimal diffeomorphism generated by $T$. 
	
	An $G$-equivariant differential form $\alpha$ is a map $\alpha : \mathfrak{g} \to \Omega^\bullet (M)$, such that for $\forall T \in \mathfrak{g}$, $\alpha(T)$ is a differential form and satisfies the relation
	\begin{equation}
		{g^*}\left( {\alpha \left( T \right)} \right) = \alpha \left( {Ad{_g}T} \right).
	\end{equation}
Let us denote the space of $G$-equivariant differential forms on $M$ by $\Omega^\bullet _G (M)$.
	
	Note that one can define the equivariant differential $d_G$ on $\Omega^\bullet _G (M)$ as
	\begin{equation}
		\left( {{d_G}\alpha } \right)\left( {{T}} \right) = d\left( {\alpha \left( T \right)} \right) - \iota \left( {{T_M}} \right)\left( {\alpha \left( T \right)} \right), \;\; \alpha \in \Omega^\bullet_G(M).
	\end{equation}
which squares to
	\begin{equation}
		\left( {{d_G}{d_G}\alpha } \right)\left( T \right) =  - {\mathcal{L}_{{T_M}}}\left[ {\alpha \left( T \right)} \right] = 0, \;\; \alpha \in \Omega^\bullet_G(M),
	\end{equation}
where the last equality uses $\alpha$ is equivariant\footnote{If $\alpha$ is a map from $\mathfrak{g} \to \alpha_M$ where $\alpha_M$ is one given usual differential form on $M$, then $\alpha_M$ being equivariant means invariant under $G$-action. In this sense, all $G$-invariant usual forms are $G$-equivariant. The action of $d_G$ on such a form $\alpha$ reads
	\begin{equation}
		\left( {{d_G}\alpha } \right)\left( X \right) = d\alpha  - {\iota _{{X_M}}}\alpha,
	\end{equation}
	}.
	
	Let us consider the differential form
	\begin{equation}
		\mathcal{J}\left( {\kappa} \right) = \kappa  \wedge \delta \left( {{d_G}\kappa } \right).
	\end{equation}
Here the $\delta$ function is viewed as a formal integral
	\begin{equation}
		\delta \left( x \right) = \frac{1}{{2\pi }}\int_{ - \infty }^{ + \infty } {{e^{itx}}dt} ,
	\end{equation}
it can take a differential form as argument, and produce a distributional differential form. Note that $\kappa$ is invariant under $G$-action, and therefore corresponds to a $G$-equivariant form.
	
	It is easy to see that $\mathcal{J}(\kappa)$ is $d_G$-closed.
	\begin{equation}
		{d_G}\left( {\kappa  \wedge \delta \left( {{d_G}\kappa } \right)} \right) = {d_G}\kappa  \wedge \delta \left( {{d_G}\kappa } \right) = 0,
	\end{equation}
where we have used $d_G^2 \kappa = 0$ and $x\delta(x) = 0$.
	
	Now suppose there is a $G$-equivariant vector bundle $V$ over $M$, $G$-invariant connection $d_A$. Then one can define its equivariant curvature $F^G_A$ as a map $F^G_A : \mathfrak{g} \to \Omega^\bullet (M, End(V))$
	\begin{equation}
		F_A^G\left( T \right)\sigma  = {F_A}\sigma  - {\mathcal{L}_{{T_M}}}\sigma  + \nabla_{T_M} \sigma , \;\; \sigma \in \Gamma(M, V).
	\end{equation}
Note that $F_A^G$ is still a tensor. Using the equivariant version of curvature, one can define all kinds of equivariant characteristic classes. The most important ones for us are the equivariant Todd class:
	\begin{equation}
		Td\left( V \right)\left( T \right) \equiv {{\det} _\mathbb{C}}\left( {\frac{{ - F_A^G\left( T \right)}}{{1 - {e^{F_A^G\left( T \right)}}}}} \right),
	\end{equation}
and equivariant Chern character:
	\begin{equation}
		Ch\left( V \right)\left( T \right) \equiv {\text{tr}}\exp \left( {\frac{1}{{2\pi i}}F_A^G\left( T \right)} \right).
	\end{equation}
	
	\vspace{20pt}
	Now we are ready to express $\dim \mathcal{M}_\lambda$ in terms of an integral over $\mathcal{M}$. As we already discussed, $\dim \mathcal{M}$ corresponds to the zeroth order term of the equivariant index of $\slashed{D_A}$, which can be identified with a Kohn-Dirac operator on K-contact manifold $(M, \kappa, R, g)$. The index of the Dirac operator has been computed in various literatures \cite{BV}\cite{Fitzpatrick:2007aa}, and in particular we now state the result from \cite{Fitzpatrick:2007aa}.
	
	Let $(M, \kappa, R)$ be a contact 5-manifold with $G$ acting elliptically. Let $\slashed{D}^+_V$ be $G$-transversally elliptic Dirac operator defined on the canonical ${\rm Spin}^\mathbb{C}$-bundle $S$ (twisted by $G$-equivariant vector bundle $V$ with connection A) via a $G$-invariant connection $\nabla$ on $S$. Then the equivariant index ${\text{in}}{{\text{d}}_G}\slashed{D}^+_V$ is a generalized function defined on $G$, and in particular, when $\exp(T)$ is close enough to the identity $e \in G$, one has\footnote{The one that is direct relevant to our case is ${\rm ind}\slashed{D}^-_V = - {\rm ind}\slashed{D}^+_V$}
	\begin{equation}
		{\text{ind}}{_G}\slashed{D}^+_V\left( {{e^T}} \right) = {\left( {\frac{1}{{2\pi i}}} \right)^2}\int_M {{\rm Td}\left( T^{0,1}_HM  \right)\left( T \right) \wedge \mathcal{J}\left( \kappa  \right)\left( T \right) \wedge Ch(V, A)(T)} ,
	\end{equation}
where ${\rm Td}(T^{0,1}_HM)$ is computed using $\nabla$.

	We start with the simplest case when $(M, \kappa, R, g)$ forms a K-contact manifold which is regular. The regularity implies $R$ generates free $G = U(1)$ action on $M$, which obvious acts elliptically. In particular, $M$ is a principal $U(1)$-bundle $\pi_{M_4}: M \to M_4$ over some symplectic base manifold $(M_4, \omega)$. 
		
	In such construction, $T_HM = \pi_{M_4}^* TM_4$. Let $S_{M_4}$ be the canonical ${\rm Spin}^\mathbb{C}$-bundle over $M_4$(see Appendix), and $\slashed{D}^+_{M_4}$ be the resulting Dolbeault-Dirac. Then $\slashed{D}_{M_4}$ pulls back to $\slashed{D}^+$ on $M$. If $S_{M_4}$ is twisted by a vector bundle $V$ on $M_4$, we can also pull it back to $M$ and form $S \otimes V$, together with a twisted Dirac operator $\slashed{D}_V^+$.
	
	Then we have a relation between the equivariant indices of two Dirac type operator:
	\begin{equation}
		{\text{in}}{{\text{d}}_{U\left( 1 \right)}}D_V^ + \left( {{e^{i\varphi }}} \right) = \sum\limits_{m \in \mathbb{Z}} {{e^{im\varphi }}{\text{in}}{{\text{d}}}D_{V \otimes {L^m}}^+} ,
	\end{equation}
where $L \to M_4$ is the associated complex line bundle of the $U(1)$-principal bundle $M \to M_4$. Notice that the $\varphi$ independent term is ${\rm ind}\slashed{D}_{V}^+$, which computes the dimension of (anti)self-dual instanton moduli on the base manifold $M_4$. Note that this is consistent with the fact that in the regular case, contact-intantons pushes down to (anti)self-dual instanton \cite{Kallen:2012aa}.

	In more general K-contact manifold, the elliptical $G$-action is the isometric $T^k$-action which is the closure of $R$-translation. Since at each point, there is always part of the $T^k$ is generated by $R$, which is transverse to $T_HM$, and therefore the $T^k$-action is elliptic. For $\left( {{e^{i{\varphi _1}}},...,{e^{i{\varphi _k}}}} \right) \in {T^k}$ near the unit $(1,...,1)$,
	\begin{equation}
		{\text{in}}{{\text{d}}_{{T^k}}}D_V^ + \left( {{e^{i{\varphi _1}}},...,{e^{i{\varphi _k}}}} \right) = {\left( {\frac{1}{{2\pi i}}} \right)^2}\int_M {{\rm Td}\left( {T_H^{0,1}M} \right)\left( {{\varphi _i}} \right) \wedge \mathcal{J}\left( \kappa  \right)\left( {{\varphi _i}} \right) \wedge Ch(V,A)\left( {{\varphi _i}} \right)} ,
	\end{equation}
where
	\begin{equation}
		\mathcal{J}\left( \kappa  \right)\left( {{\varphi _i}} \right) = \frac{1}{{2\pi }}\sum\limits_{m \in \mathbb{Z}} {\left( {\kappa  + imd\kappa  - {m^2}{{\left( {d\kappa } \right)}^2}} \right){e^{ - im\sum\limits_i {{\varphi _i}\kappa \left( {{T_i}} \right)} }}} ,
	\end{equation}
where $\{T_i\}$ are the vector fields on $M$ that correspond to the basis of Lie algebra $\mathfrak{t}^k$. The $\varphi_i$ independent term of the integral then gives the $\dim \mathcal{M}$.

	\section{5d $\mathcal{N} = 1$ Yang-Mills and 6d Donaldson-Thomas theory}
	
	This section separates into two parts. First we will review the relation of the cohomological theory (\ref{Lagrangian}) with $\mathcal{N} = 1$ super-Yang-Mills in 5-dimension, and using a canonical contact-instanton solution we discuss how the BRST Wilson loop operator $W(\gamma)$ is related to the linearized return map $\Psi_\gamma$ appears in contact homology \cite{Contact-Homology}. Second, we will discuss an induced 6d cohomological theory, which relates to Donaldson-Uhlenbeck-Yau equation.

	\subsection{5d $\mathcal{N} = 1$ Yang-Mills and linearized return map}
	
	The cohomological field theory defined earlier is closely related but not identical to $5d$ $\mathcal{N} = 1$ super Yang-Mills theory considered in \cite{Kallen:2012ab}.
	
	The $\mathcal{N} = 1$ vector multiplet consists of the following field contents:
	consists of the following field content:
	\begin{itemize}
		\item Hermitian gauge field $A_m$; covariant derivative $D_m = \nabla_m - i [A, \bullet]$ and hermitian field strength $F = dA -i A\wedge A$;
		\item Real scalar $\phi$;
		\item Spinor $\lambda_I$, satisfying symplectic-Majorana condition $\overline {\lambda _I^\alpha }  = {\epsilon ^{IJ}}{C_{\alpha \beta }}\lambda _J^\beta $. Note that it has 8 real degrees of freedom;
		\item Auxiliary scalar field $D_{IJ}$, with reality condition $\overline{D_{IJ}} = \epsilon^{II'}\epsilon^{JJ'} D_{I'J'}$.
	\end{itemize}

	The supersymmetry transformation defined on $\mathbb{R}^5$ is (with $SU(2)$ symplectic Majorana spinor $\xi_I$ as parameter):
	\begin{equation}
		\left\{ \begin{array}{l}
		{\delta _\xi }{A_m} = i{\epsilon ^{IJ}}{\xi _I}{\Gamma _m}{\lambda _J}\\[0.5em]
		{\delta _\xi }\phi  = i{\epsilon ^{IJ}}{\xi _I}{\lambda _J}\\[0.5em]
		\displaystyle {\delta _\xi }{\lambda _I} =  - \frac{1}{2}{F_{mn}}{\Gamma ^{mn}}{\xi _I} + \left( {{D_m}\phi } \right){\Gamma ^m}{\xi _I} + {\epsilon ^{JK}}{\xi _J}{D_{KI}}\\[0.5em]
		{\delta _\xi }{D_{IJ}} =  - i{\xi _I}{\Gamma ^m}{D_m}{\lambda _J} + \left[ {\sigma ,{\xi _I}{\lambda _J}} \right] + \left( {I \leftrightarrow J} \right)
		\end{array} \right..
	\end{equation}
	
	Given any even $SU(2)$-symplectic Majarana spinor $\xi_I$, one can redefine $\lambda_I$ and $D_{IJ}$ in terms of odd 1-form $\psi$, odd 2-form $\chi$ and even 2-form $H$:
	\begin{equation}
		\left\{ \begin{array}{l}
		{\psi _m} \equiv {\epsilon ^{IJ}}\left( {{\xi _I}{\Gamma _m}{\lambda _J}} \right)\\[0.5em]
		{\chi _{mn}} \equiv  {\epsilon ^{IJ}}\left[ {\left( {{\xi _I}{\Gamma _{mn}}{\lambda _J}} \right) - {\kappa _m}\left( {{\xi _I}{\Gamma _n}{\lambda _J}} \right) + {\kappa _n}\left( {{\xi _I}{\Gamma _n}{\lambda _J}} \right)} \right]\\[0.5em]
		H \equiv 2 F_H^ -  + {D^{IJ}}{\Theta _{IJ}}
		\end{array} \right..
	\end{equation}
where ${\left( {{\Theta _{IJ}}} \right)_{mn}} = \left( {{\xi _I}{\Gamma _{mn}}{\xi _J}} \right)$. Note that $\chi$ and $H$ so defined satisfy the self-dual property
	\begin{equation}
		{\iota _R}\chi  = {\iota _R}H = 0,\;\;{\iota _R}*\chi  = \chi ,\;\;{\iota _R}*H = \chi ,
	\end{equation}
with normalization $(\xi_I \xi^I) = 1$ and vector field $R^m \equiv - (\xi_I \Gamma^m \xi^I)$\footnote{Note the minus sign; different signs will flip the self-duality property.}.

	After the redefinition, the supersymmetry transformation can be rewritten as \cite{Kallen:2012aa}
	\begin{equation}
		\left\{ \begin{array}{l}
		{\delta }{A} = i {\psi}\\[0.5em]
		{\delta }\phi  = - {\iota _R}\psi \\[0.5em]
		\delta {\psi } = i {\iota _R}F + {d_A}\phi \\[0.5em]
		\displaystyle \delta \chi  = H \\[0.5em]
		\delta H = -\mathcal{L}_R^A\chi  - \left[ {\phi ,\chi } \right]
		\end{array} \right..
		\label{delta-transformation}
	\end{equation}
	
	The supersymmetry can be deformed and generalized to curve manifold. In \cite{Hosomichi:2012aa}, supersymmetry is defined on $S^5$ with Killing spinor $\xi_I$
	\begin{equation}
		{\nabla _m}{\xi _I} = {\Gamma _m}{t_I}^J{\xi _J}.
	\end{equation}
This is further studied in \cite{Kallen:2012ab}, where the fields are redefined as above. The $\delta$-transformation (\ref{delta-transformation}) and corresponding invariant theory can be defined on any K-contact manifolds. In \cite{Pan:2013aa}, a generalized supersymmetry and supersymmetric theory for $\mathcal{N} =1 $ vector multiplet is proposed. Under the field redefinition, the $\delta$-transformation of $\chi$ is modified:
	\begin{equation}
		\delta \chi  = H - 2\lambda \phi d\kappa  + \Omega _H^ +,
	\end{equation}
where $\Omega_H^+$ and $d\kappa$ are both self-dual, with the transformation of other fields identical to (\ref{delta-transformation}). 

	It is easy to see that part of our multiplet is equivalent to the 5d $\mathcal{N} = 1$ vector multiplet, with additional $\chi_V$, $H_V$, $\bar \phi$ and $\eta$ as contactable pairs. The $Q$-transformation can be naively considered as the $\delta$-transformation with $R = 0$ and $\Omega_H^+ = 0$.
	
	let us consider modifying (\ref{delta-transformation}) such that
	\begin{equation}
		\begin{gathered}
	 	 \delta \chi _H^ +  = H_H^ +  - 2\lambda \phi d\kappa  \hfill \\[0.5em]
	 	 \delta H_H^ +  =  - \mathcal{L}_R^A\chi _H^ +  - \left[ {\phi ,\chi _H^ + } \right] - 2\lambda ({\iota _R}\psi) d\kappa  \hfill \\ 
		\end{gathered} ,
	\end{equation} 
to impose global condition $d\kappa \ne 0$ naturally. Operator $\delta$ squares to translation along $R$ and gauge transformation:
	\begin{equation}
		\delta^2 = -\mathcal{L}_R + iG_{A+i\phi}.
	\end{equation}
One can define a twisted $\mathcal{N} = 1$ theory $\mathcal{L}_{\mathcal{N} = 1} = \delta V$ on any K-contact manifold $M$ similar to  that in \cite{Kallen:2012ab}:
	\begin{equation}
		V = \frac{1}{{{e^2}}}\delta \left( {\frac{1}{2}\chi  \wedge *\left( {2F - H} \right) + \psi  \wedge *\overline {\delta \psi } } \right),
	\end{equation}
which computes homotopy invariants of K-contact structures. The $\delta$-invariant observables are Chern-Simons type observables, and Wilson loops along integral loops $\gamma$ of the Reeb vector field (if exist):
	\begin{equation}
		I\left( {\gamma ,k} \right) \equiv {\text{tr}}\left[ {{W }{{\left( \gamma \right)}^k}} \right],\;\;{W }\left( \gamma \right) \equiv P\exp \left[ {\oint_\gamma  {A + i\phi \kappa } } \right] = hol_\gamma(A+i\phi\kappa).
	\end{equation}
	
	It is straight forward to see that the theory defined is invariant up to $\delta$-exact terms under deformation $(\Delta \kappa, \Delta R, \Delta g)$, as long as they preserve K-contact conditions. In particular, the expectation values $\left\langle I\left( {\gamma ,k} \right) \right\rangle _\mathfrak{R} $  are homotopy invariants of of K-contact structures, where $\mathfrak{R}$ denotes some irreducible representation of the gauge group. Note that if $A$ is irreducible, $W_\gamma (A)$ is just the holonomy of $A$ along $\gamma$, since $\phi = 0$ by irreducibility.
	
	At the end of appendix [\ref{Appendix-B}], we discussed an interesting property of the generalized Tanaka-Webster connection $\nabla$ on $E \equiv T_H M$, namely its holonomy $W(\gamma, \nabla)$ along closed Reeb integral curves $\gamma$ coincides with the linearized return map $\Psi_{\gamma}$. Linearized return map is defined as the map ${\psi _p}:{E_p} \to {E_p}$ as the restriction of the Reeb flow, where $p$ is a point in the closed Reeb curve $\gamma$, 
	
	Now let us consider the twisted $\mathcal{N} = 1$ theory with $G = U(2)$ and consider $E = T_HM$ as the $G$-bundle. Namely, a canonical twisted $\mathcal{N} = 1$ theory associated to the K-contact structure. It would be very interesting if two things happen: 1) $\nabla$ is actually a contact-instanton on $E$ as a $G = U(2)$-vector bundle, onto which the field theory localizes, and 2) the expectation values of Wilson loops give information about the linearized return map $\Psi_\gamma$\cite{Contact-Homology}.
	
	Although it is not clear if this can be achieved in general, however, on any Sasaki-Einstein manifold $M$ this is indeed the case. In \cite{Harland:2011aa}, another version of canonical connection $\nabla^P$ on $TM$ is defined 
	\begin{equation}
		\left\{ \begin{gathered}
	 	^P{\Gamma ^b}_{ma} = {{\dot \Gamma }^b}{\;_{ma}} + \frac{1}{2}{\left( {\frac{1}{2}\kappa  \wedge d\kappa } \right)_{mab}} \hfill \\
  		-^P{\Gamma^5}_{ma} = ^P{\Gamma ^a}_{m5} = {{\dot \Gamma }^a}{\;_{m5}} + {\left( {\frac{1}{2}\kappa  \wedge d\kappa } \right)_{m5a}} \hfill \\ 
		\end{gathered}  \right..
	\end{equation}
where $^P\Gamma$ denotes the connection coefficients of $\nabla^P$. The significance of $\nabla^P$ is that it preserves $\kappa$ and $\Phi$, and its curvature $F_{\nabla^P}$ satisfies
	\begin{equation}
		\pi _H^ + {F_{{\nabla ^P}}} = 0,\;\;\;\;{\pi _V}{F_{{\nabla ^P}}} = 0.
	\end{equation}
Viewed as connections on $E$, $\nabla^P$ and generalized Tanaka-Webster $\nabla$ differ by
	\begin{equation}
		{\nabla _m} = \nabla _m^P - \frac{1}{4}\Phi {\kappa _m},
	\end{equation}
and therefore the curvatures differ by
	\begin{equation}
		{F_\nabla } = {F_{{\nabla ^P}}} - \frac{1}{4}\Phi d\kappa,
	\end{equation}
where $\nabla^P \Phi = 0$ has been used. Clearly, we can read off from this equation
	\begin{equation}
		{\iota _R}{F_\nabla } = 0,\;\;\;\;\pi _H^ + {F_\nabla } =  - \frac{1}{4}\Phi d\kappa ,\;\;\;\; {d_\nabla }\Phi  = 0,
	\end{equation}
namely, the generalized Tanaka-Webster connection $\nabla$ is a reducible contact-instanton. It also satisfies the gauge condition
	\begin{equation}
		{\mathcal{L}_R}\Gamma  = 0,
	\end{equation}
following from K-contact conditions $\mathcal{L}_R g = \mathcal{L}_R \Phi = 0$. Finally, noting the fact that $\Phi$ is invariant under parallel transport of $\nabla$, we have
	\begin{equation}
		P\exp \left[ {\oint_\gamma  {\Gamma  + \Phi \kappa } } \right] = ho{l_\gamma }\left( {\nabla  + \Phi \kappa } \right) = ho{l_\gamma }\left( \nabla  \right) = {\Psi _\gamma }.
	\end{equation}
	
	Wilson loop does not provide enough information for us to explore further. To understand Conley-Zehnder index and get concrete relation to contact homology from gauge theory perspective, one may need to incorporate hypermultiplet and study supersymmetric line operators that connect ``crossing points"\cite{Contact-Homology}. We leave this to future study.

	\vspace{10pt}
	\subsection{6d cohomological theory and Donaldson-Uhlenbeck-Yau equation}
	
	The cohomological theory can also be uplifted to a 6d gauge theory which localizes to interesting configurations.
	
	In \cite{L.Baulieu:aa}, for each $d$-dimensional cohomological field theory (called $\mathcal{H}_d$), a $d + 1$-dimensional theory $\mathcal{K}_d$ is constructed. Using that method, we can define for our $5d$ theory a corresponding $6d$ theory. Before discussing the $6d$ theory, let us first introduce a 6-dimensional symplectic manifold $X$ on which we will put a theory, associated to the 5-dimensional contact manifold $M$.
	
	Consider $X = M \times \mathbb{R}$, with $r > 0$ as the coordinate along $\mathbb{R}$. Accordingly, the tangent bundle
	\begin{equation}
		TX = TM \oplus T\mathbb{R} = T_H M \oplus \mathbb{R} R \oplus T\mathbb{R}.
	\end{equation}
Denote the projection $\pi_M : X \to M$, which maps $\pi_M(p, r) = p$.

	Let us define a closed 2-form $\omega \equiv \frac{1}{2}d(r^2 \kappa) = rdr \wedge \kappa + \frac{1}{2}r^2 d\kappa$. It is obvious that it defines a symplectic structure on $X$:
	\begin{equation}
		\omega ^3  = \frac{1}{8}\left( {2rdr \wedge \kappa  + r^2 d\kappa } \right)^3  = \frac{1}{4}r^5 dr \wedge \kappa  \wedge \left( {d\kappa } \right)^2  \ne 0,
	\end{equation}
where the term $(d\kappa)^3 = 0$ since $d\kappa$ vanishes on $R$. There is another coordinate $t$ on $\mathbb{R}$ that is frequently used:
	\begin{equation}
		e^t  \equiv r^2 .
	\end{equation}
In this coordinate, $\omega  = e^t \left( {dt \wedge \kappa  + d\kappa } \right)$. One can also define a metric on $X$ via
	\begin{equation}
		g_X  = dr^2  + r^2 g_M ,
	\end{equation}
where $g_M$ is tha associated metric of $\kappa$ on $M$. Similarly, one cah extend the tensor $\Phi$ to an almost complex structure $J$ on $X$, by defining
	\begin{equation}
		\left. J \right|_{T_H M}  = \Phi ,\;\;J\left( R \right) =  - r\partial _r \equiv - H.
	\end{equation}
Here the name $H$ comes from "Homothety"; note that $J^2 = -1$ on the tangent space $TX$. As usual, $J$ provides a complex decomposition of $TX_\mathbb{C}$ and a $(p,q)$-decomposition of $\wedge^\bullet TX^*_\mathbb{C}$. What is most useful for us, is that the original decompositions on $M$ automatically fits in the $6d$ picture. For instance, any real $F \in \Omega^1(X)$ can be decomposed into
	\begin{equation}
		A  = A ^{1,0}  + A ^{0,1}  = A _H^{1,0}  + a\left( {r^{ - 1} dr + i\kappa } \right) + A _H^{0,1}  + \bar a\left( {r^{ - 1} dr - i\kappa } \right),
	\end{equation}
where $A^{1,0} \in \Gamma(\pi_M^* T^{1,0}M^*)$ and $A^{1,0} = \overline{A^{0,1}}$, $a \in C^\infty(X, \mathbb{C})$.

	Similarly, any real 2-form $F$ on $X$ can be decomposed into
	\begin{equation}
		\begin{gathered}
  		F  = F _H^{2,0}  + \left( {r^{ - 1} dr + i\kappa } \right) \wedge \mu _H^{1,0}  + F _H^{0,2}  + \left( {r^{ - 1} dr - i\kappa } \right) \wedge  {\mu _H^{0,1} }  \hfill \\[0.5em]
  		\;\;\;\;\;\;\; + \xi _H^{1,0}  \wedge \chi _H^{0,1}  + \frac{1}{4} l  d\kappa  + br^{ - 1} \kappa  \wedge dr + \nu _H^{1,0}  \wedge \left( {r^{ - 1} dr - i\kappa } \right) + c.c. \hfill, \\
		\label{6d-decomposition}
		\end{gathered} 
	\end{equation}
where the first row corresponds to the $(2,0)$ and $(0,2)$ components, while the second row corresponds to $(1,1)$ components, and $F^{2,0} = \overline{F^{0,2}}$, $\mu_H^{1,0} = \overline{\mu_H^{0,1}}$, $l, b \in C^\infty (M, \mathbb{R})$. Notice also that the three terms
	\begin{equation}
		F _H^{2,0}  + F _H^{0,2}  + ad\kappa  = F _H^ +  ,
	\end{equation}
forms the horizontal-self-dual part of $F$, while
	\begin{equation}
		\xi_H^{1,0} \wedge \zeta_H^{0,1} + c.c. = F_H^-	,
	\end{equation}
forms the horizontal anti-self-dual part of $F$. 	
	
	The projection $\pi : X \to M$ pulls back the gauge bundle over $M$, which we continue to denote as $P_G$ and ${\rm ad}P_G$. We consider the field contents
	\begin{equation}
		\{A,\;\;\psi ,\;\;\sigma ,\;\;\chi ,\;\;H \},
	\end{equation}
where $A$ is $6d$ connection, and $\psi$ is a $6d$ 1-form, $\sigma$ is a real scalar, and $\chi$ and $H$ corresponds to the pull back of original $\chi$ and $H$.

	Following \cite{L.Baulieu:aa}, we define the transformation $\delta$ as
	\begin{equation}
		\left\{ \begin{gathered}
  		\left\{ {\delta ,A} \right\} = i\psi  \hfill \\[0.5em]
  		\left\{ {\delta ,\psi } \right\} = i\iota _{\partial /\partial t} F + d_A \sigma  \hfill \\[0.5em]
  		\left\{ {\delta ,\sigma } \right\} =  - \psi _t  \hfill \\[0.5em]
  		\left\{ {\delta ,\chi_H^+ } \right\} = H_H^+ + 2\lambda \sigma d\kappa,\;\;\;\;\;\;\;\;\;\;\;\;\;\;\;\;\;\;\;\;\;\;\;\;\;\;\;\;\;\;\;\;\; \{\delta, \chi_V\} = H_V \hfill \\[0.5em]
  		\left\{ {\delta ,H_H^ + } \right\} =  - \partial _t^A\chi _H^ +  - \left[ {\sigma ,\chi _H^ + } \right] + 2\lambda {\psi _t}d\kappa ,\;\;\;\;\;\;\;\;\;\{ \delta ,{H_V}\}  = {\chi _V}  \hfill \\ 
		\end{gathered}  \right..
	\end{equation}
Transformation $\delta$ squares to translation along $\mathbb{R}$ and gauge transformation:
	\begin{equation}
		\delta ^2 = -\partial_t + i\mathcal{G}_{A_t + i\sigma}.
	\end{equation}
	
	The Lagrangian in $6d$ is a straightforward generalization of (\ref{Lagrangian}):
	\begin{equation}
		\mathcal{L}_{d = 6}  = \left\{ {\delta ,\frac{1}{{e^2 }}\int_M {{\text{tr}}\left[ {\frac{1}{2} \pi^- \chi  \wedge *\left( {2F - H} \right) + \left( {d_A \sigma  + \iota _{\partial /\partial t} F} \right) \wedge *\psi } \right]} } \right\}.
		\label{6d-theory}
	\end{equation}
It is then easy show that the partition function localizes to solitonic solutions
	\begin{equation}
		\iota _{\partial /\partial t} F = 0,\;\;F_H^ +   = \lambda \sigma d\kappa ,\;\;F_V  = 0,\;\;d_A \sigma  = 0.
		\label{6d-locus}
	\end{equation}
	
	Using the decomposition (\ref{6d-decomposition}), we see that the above localization locus satisfies equations
	\begin{equation}
		{F^{2,0}} = {F^{0,2}} = 0,\;\;F \wedge \omega  \wedge \omega  = l \omega  \wedge \omega  \wedge \omega, \;\; d_A \sigma = 0.
	\end{equation}
Note that, when $M$ is Sasaki-Einstein, and therefore $X$ is Calabi-Yau, the above equations is the almost the same as Donaldson-Uhlenbeck-Yau equation\footnote{Note however that (\ref{6d-locus}) does not represent all the solutions to the DUY equation, as the latter allows non-zero $b$ and $\nu_H^{1,0}$.}, with the difference being we have a real scalar instead of complex.
	
	The theory (\ref{6d-theory}) also has two different types of observables, including the Wilson loop 
	\begin{equation}
		\mathcal{O}_\gamma ^{\left( k \right)} \equiv {\text{tr}}\left[ {W{{\left( \gamma  \right)}^k}} \right],\;\;\;\;W\left( \gamma  \right) \equiv P\exp \oint_\gamma  {\left( {{A_t} + i\sigma } \right)}, 
	\end{equation}
and the obervables of Chern-Simons type.

	\section{Summary}
	
	We introduced a notion of contact-instanton, as a generalization of that in \cite{Kallen:2012aa}, and discussed its basic properties. A cohomological field theory is proposed, whose partition function and expectation values localize to contact-instanton configurations. Then by standard arguments, we see that these quantities are, at least formally, invariant under homotopy of contact structures, or equivalently, they computes contact invariants.
	
	As the first step to understanding the path integral, we need to know the dimension of moduli space $\mathcal{M}$. However, although the cohomological theory and contact-instantons are defined for any contact structure, the deformation problem of contact-instanton turns out to be non-elliptic. At such point, we focus on a special while still vast enough class of contact structures, namely K-contact structures. On these structures, the deformation problem is recast into a transversally elliptic one, and the relevant transversally elliptic operator is shown to be the same as a Kohn-Dirac operator well-defined on any contact metric structure, whose equivariant index has been computed previously.
	
	In the comparison with 5d $\mathcal{N} = 1$ super-Yang-Mills in 5d, we slightly modify the theory in \cite{Kallen:2012ab}, and point out the connection between Wilson loops observables $W(\gamma)$ along closed integral curves $\gamma$ of $R$ and contact homology, by studying the generalized Tanaka-Webster connection.
	
	One of the remaining puzzles is the gauge choice that we made to reformulate the deformation problem, namely
	\begin{equation}
		{\mathcal{L}_R}A = 0.
	\end{equation}
It is not clear to the author how general this choice could be, and in fact, we believe that such choice is only valid in certain sector of the space $\mathcal{A}$ of connections. For instance, combining with irreducibility, this would implies $\iota_R A = 0$, which is not true if $A$ has non-trivial holonomy along any closed integral curve of $R$, if there is any\footnote{Any closed 3-dimensional contact manifold must have at least one closed Reeb integral curve, proven by Taubes.}. Actually,  from our current result that $\dim \mathcal{M}$ is given by the zeroth order term of ${\rm ind}\slashed{D}^-_{S^+\otimes {\rm ad}P_G}$, we can guess that the compete answer should have contributions from all the terms, corresponding to different sectors (and therefore different admissible gauge choices) where $A$ has different behaviors along $R$. Note that these gauge choices should enable reformulations of (\ref{original_deformation}) into something more tractable, like (\ref{reduced-deformation-complex}).
	
	We used the Killing condition $\mathcal{L}_R g = 0$ to construct a canonical elliptic torus-action and establish a vanishing theorem. It is not clear to the author if the Killing condition can be dropped, and generalize the result to generic contact structures. 
	
	One may want to work entirely with K-contact structure instead. One can directly use the $\mathcal{N} = 1$ supersymmetry in its twisted form as discussed in the previous section and \cite{Kallen:2012ab}, and study the equivariant intersection theory coming out from the resulting cohomological theory. In this way one may get interesting information about K-contact structures from the expectation values of the observables, in particular, the Wilson loop observables introduced earlier, which encode information about closed integral curves of the Reeb vector fields.
	
	Therefore, as a canonical application, it will be interesting to explicitly compute the VEV of Wilson loops of the twisted $\mathcal{N} = 1$ theory with $G = U(2)$ and the $G$-bundle as $E = T_HM$,
	\begin{equation}
		\left\langle {\prod\limits_i {I\left( {{\gamma _i},{k_i}} \right)} } \right\rangle  =  {\prod\limits_i {{\text{tr}}{{\left( {{\Psi _{{\gamma _i}}}} \right)}^{{k_i}}}} }   +  {...} 
	\end{equation}
which could be invariant under homotopy of K-contact structures. More interesting quantities requires coupling to hypermultiplets and study Wilson lines that connects quarks inserted at special points on closed Reeb curves.

Finally, the reducible contact-instantons need a more careful treatment. We already come across such a solution, namely the generalized Tanaka-Webster connection on the $U(2)$-bundle $T_HM$. In that case, we show that it contains interesting information about the underying K-contact structure. It is natural to expect that most information about the underlying geometries lie in the reducibles, since the main difference between the $\lambda = 0$ and $\lambda \ne 0$ version of (\ref{contact-instanton}) concentrates on reducibles.

\section*{Acknowledgments}

The author thanks Sean Fitzpatrick for discussions on related mathematics. The author also would like thank NSF grant no. PHY-1316617 for partial support.

	\begin{appendices}
	
	\section{Differential Geometry and Notations}\label{Appendix-A}
	
	In this appendix, we fix some of the notations and briefly summarize relevant formulae in differential geometry.
	
	\vspace{20pt}
	
	For a smooth manifold $M$, we denote $TM$ as its tangent bundle, and its fiber at $p \in M$ is denoted as $TM_p$. The cotangent bundle is denoted as $TM^*$, and the bundle of $p$-forms is denoted as $\wedge^p TM^*$ with the space of sections denoted as $\Omega^p(M)$. The space of Lie-algebra $\mathfrak{g}$-valued $p$-forms is denoted as $\Omega^p(M, \mathfrak{g})$.
	
	Any differential $p$-form $\omega$ can be written using components $\omega_{m_1...m_p}$
	\begin{equation}
		\omega  = \frac{1}{{p!}}{\omega _{{m_1}...{m_p}}}d{x^{{m_1}}} \wedge ... \wedge d{x^{{m_p}}},
	\end{equation}
where $\omega_{m_1...m_p}$ is totally anti-symmetric.
	
	Let $X$ be a vector field and $\omega$ be a $p$-form. The contraction of $X$ and $\omega$ is a $p-1$-form, and denoted as $\iota_X \omega$, or in components
	\begin{equation}
		{\left( {{\iota _X}\omega } \right)_{{m_1}...{m_{p - 1}}}} = {X^n}{\omega _{n{m_1}...{m_{p - 1}}}}.
	\end{equation}
	
	The exterior derivative $d$ acts on $\omega$ as
	\begin{equation}
		d\omega  = \frac{1}{{p!}}{\partial _k}{\omega _{{m_1}...{m_p}}}d{x^k} \wedge d{x^{{m_1}}} \wedge ... \wedge d{x^{{m_p}}},
	\end{equation}
and in particular, when $\omega$ is a 1-form or 2-form, the action can be easily written in components:
	\begin{equation}
		{\left( {d\omega } \right)_{mn}} = {\partial _m}{\omega _n} - {\partial _n}{\omega _m},\;\;\;\;{\left( {d\omega } \right)_{kmn}} = {\partial _m}{\omega _{nk}} + {\partial _n}{\omega _{km}} + {\partial _k}{\omega _{mn}}.
	\end{equation}
	
	Given a Riemannian metric $g$, one can define the adjoint $d^*$ of $d$: when acting on a $p$-form,
	\begin{equation}
		{d^*} = {\left( { - 1} \right)^{np + n + 1}}*{d} \;*,
	\end{equation}
or in components
	\begin{equation}
		{\left( {{d^*}\omega } \right)_{{m_1}...{m_{p - 1}}}} =  - {{\dot \nabla }^n}{\omega _{n{m_1}...{m_{p - 1}}}}	,
	\end{equation}
where $\dot \nabla$ denotes the Levi-civita connection of $g$.

	Using $\iota_X$ and $d$, the Lie-derivative with respect to $X$ is defined by Cartan's formula
	\begin{equation}
		{\mathcal{L}_X}\omega  = \left( {d{\iota _X} + {\iota _X}d} \right)\omega, \;\;\;\; \forall \omega \in \Omega^\bullet (M).
	\end{equation}
	
	The Lie-derivative can also acts on arbitrary tensors. When acting on a vector field $Y$,
	\begin{equation}
		{\mathcal{L}_X}Y = \left[ {X,Y} \right],
	\end{equation}
while acting on $(1,1)$-type tensor $\Phi$ and $(0,2)$-type tensor $G$,
	\begin{equation}
		{\left( {{\mathcal{L}_X}\Phi } \right)^m}_n = {X^k}{\partial _k}{\Phi ^m}_n - \left( {{\partial _k}{X^m}} \right){\Phi ^k}_n + \left( {{\partial _n}{X^k}} \right){\Phi ^m}_k;
	\end{equation}
	\begin{equation}
		{\left( {{\mathcal{L}_X}G} \right)_{mn}} = {X^k}{\partial _k}{G_{mn}} + \left( {{\partial _m}{X^k}} \right){G_{kn}} + \left( {{\partial _n}{X^k}} \right){G_{mk}}.
	\end{equation}
	
	If $M$ is equipped with a metric $g$ with Levi-civita connection $\dot \nabla$, then $\partial_k$ can be replaced by $\dot \nabla_k$ in all the above formula. In particular, $X$ is Killing vector field if
	\begin{equation}
		\mathcal{L}_X g = 0 \Leftrightarrow \dot \nabla_m X_n + \dot \nabla_n X_m = 0.
	\end{equation}
	
	The curvature of the Levi-civita connection of $g$ is defined as
	\begin{equation}
		\left[ {{\nabla _m},{\nabla _n}} \right]{X^k} = {R^k}_{lmn}{X^l},
	\end{equation}
and we also define the curvature tensor with all indices down ${R_{klmn}} = {g_{kk'}}{R^{k'}}_{lmn}$. This tensor satisfies various identities:
	\begin{equation}
		{R_{klmn}} = {R_{mnkl}},\;\;\;{R_{klmn}} =  - {R_{klnm}},\;\;\;{R_{klmn}} + {R_{kmnl}} + {R_{knlm}} = 0.
	\end{equation}
	
	The Ricci tensor is defined as
	\begin{equation}
		Ri{c_{mn}} \equiv \sum\limits_k {{R^k}_{mkn}} ,
	\end{equation}
which is symmetric. Finally the scalar curvature is defined as $\mathcal{R} \equiv g^{mn} Ric_{mn}$.
	
	Other than the Levi-civita connection, there are a lot of connections that preserve the metric $g$. Let $\nabla$ be arbitrary metric connection, then its connection coefficients can be written in terms of the christoffel symbol and contorsion tensor $C$:
	\begin{equation}
		{\nabla _m}{X^k} = {\partial _m}{X^k} + {\Gamma ^k}_{mn}{X^n},\;\;\;\;{\Gamma ^k}_{mn} = {\dot \Gamma ^k}{\;_{mn}} + {C^k}_{mn}.
	\end{equation}
The anti-symmetric part of $C^k_{mn}$ is the torsion tensor.

	\section{Contact and almost contact structure in $d = 5$} \label{Appendix-B}
	
	Contact geometry is the odd dimensional cousin of symplectic geometry which exists in even dimension. It is however much less studied compared with the latter. In this section we review some basics of contact geometry with emphasis on contact 5-manifolds. We refer interested readers to a beginner-friendly book \cite{Blair}, and a more involved book \cite{Geiges} for topological aspects.
	
	\vspace{20pt}
	
	\underline{Hyperplane field}

	A hyperplane field $E$ on a manifold M is a codimension one sub-bundle of the tangent bundle $TM$. Locally, $E$ can always be defined as the kernel of certain 1-form $\kappa$. In particular, any nowhere-vanishing 1-form $\kappa$ defines a global hyperplane field $E = \ker (\kappa)$. Note that rescaling $\kappa \to e^f \kappa$ does not change the corresponding hyperplane field. If $M$ is further equipped with a Riemannian metric $g$, one can define a vector field $R$ associated to $\kappa$
	\begin{equation}
		g\left( {R, \cdot } \right) \equiv \kappa \left(  \cdot  \right).
	\end{equation}
	
	\vspace{20pt}
	\underline{Almost contact structure}
	
	Let $M$ be a $2n+1$ oriented dimensional smooth manifold. An almost contact structure\footnote{An almost contact structure can also be defined as a reduction of structure group from $SO(2n+1)$ to $U(n)$.} on $M$ consists of a nowhere-vanishing 1-form $\kappa$, a nowhere vanishing vector field $R$ and a $(1,1)$-type tensor ${\Phi^m}_n$ viewed as a map $\Phi: \Gamma(TM) \to \Gamma(TM)$, such that
	\begin{equation}
		\kappa \left( R \right) = 1,\;\;{\Phi ^2} =  - 1 + R \otimes \kappa .
	\end{equation}
Note that the condition $\Phi \left( R \right) = \kappa \circ \Phi  = 0$ can be derived from the above conditions.
	
	Given an almost contact structure, one can always find a (actually infinitely many) compatible metric $g$ such that
	\begin{equation}
		g\left( {R, \cdot } \right) = \kappa \left(  \cdot  \right).
	\end{equation}
Together with the metric, $(\kappa, R, \Phi, g)$ is called an almost contact metric structure.

	\vspace{20pt}
	\underline{Contact structure, associated metric and anti-self-duality}
	
	Recall that given any nowhere-vanishing 1-form $\kappa$, one can consider its annihilating distribution ${\rm ker}( \kappa)$, namely the horizontal hyperplane field. The distribution is integrable is equivalent to the Frobenius integrability condition
	\begin{equation}
		\kappa  \wedge d\kappa  = 0.
	\end{equation}
	
	Contact structures sit on the opposite end: they correspond to completely non-integrable distributions. A nowhere-vanishing 1-form $\kappa$ defines a contact distribution ${\rm ker}(\kappa)$ if it satisfies
	\begin{equation}
		\kappa  \wedge {\left( {d\kappa } \right)^n} \ne 0.
	\end{equation}
Note that any rescaling of $\kappa \to \lambda \kappa$ with $\lambda \in C^\infty (M)$ does not change the underlying distribution, and moreover,
	\begin{equation}
		\lambda \kappa  \wedge {\left( {d\left( {\lambda \kappa } \right)} \right)^n} = \lambda \kappa  \wedge {\left( {d\lambda  \wedge \kappa  + \lambda d\kappa } \right)^n} = {\lambda ^{n + 1}}\kappa  \wedge {\left( {d\kappa } \right)^n} \ne 0.
	\end{equation}
In view of this, by contact structure we mean the distribution itself rather than the associated 1-form $\kappa$, and $\kappa$ is called the contact 1-form corresponding to the contact structure. To effectively deform a contact structure, the contact 1-form $\kappa$ must vary ``horizontally". Note that $d\kappa$ is a 2-form at each point $p \in M$ with maximal rank $2n$ by maximally non-integrable condition, and therefore there exists one vector $R_p \in T_p(M)$ such that $\iota_R d\kappa = 0$. However, $\kappa_p(R_p)$ cannot be 0 simultaneously, since
	\begin{equation}
		\kappa  \wedge {\left( {d\kappa } \right)^n}\left( R_p \right) \ne 0.
	\end{equation}
By properly rescaling, we can always choose $R_p$ such that $\kappa_p(R_p) = 1$. Do this at each point on $M$, then one obtains vector field $R \in \Gamma(TM)$, such that
	\begin{equation}
		\kappa \left( R \right) = 1,\;\;{\iota _R}d\kappa  = 0.
	\end{equation}
Then the horizontal forms are differential forms that are annihilated by $\iota_R$ and vertical forms are those of the form $\kappa \wedge (...)$. Then effective deformations of contact structure are $\delta \kappa$ such that
	\begin{equation}
		{\iota _R}\delta \kappa  = 0.
	\end{equation}
	
	Note that the integral curve of $R$ can have different types of behaviors. If the curves are all closed, then $R$ generates locally free $U(1)$-action on $M$, which is usually called "quasi-regular" contact structure. If the action is actually free, then $M$ is a principal $U(1)$-bundle over some base manifold, which is called "regular" contact structure. Otherwise, if the curves are not all closed, the contact structure is called irregular. Note however that regularity is not the intrinsic property of the contact distribution but of the contact 1-form: the regularity can be modified by rescaling $\kappa \to e^f \kappa$.
	
	Well-known example of regular contact structure is the Hopf-fibration of $S^{2n+1}$, and example of irregular contact structure can be found on $T^{2n+1}$\footnote{Note that all the contact structures on $T^{2n+1}$ are irregular. }.

	A contact structure is a special case of almost contact structure. Namely, given any contact structure, one can define an associated metric $g$ and a tensor $\Phi$ such that
	\begin{equation}
		{\Phi ^2} =  - 1 + R \otimes \kappa ,\;\;\;\;g\left( {R, \cdot } \right) = \kappa \left(  \cdot  \right),\;\;\;\;2 g\left( {X,\Phi Y} \right) = d\kappa \left( {X,Y} \right).
	\end{equation}
This set of quantities $(\kappa, R, g, \Phi)$ arising from a contact structure is called a contact metric structure. Combining the first and third equation one arrives at
	\begin{equation}
		{\left( {d\kappa } \right)_{mn}}{\left( {d\kappa } \right)^{mn}} = 16,
	\end{equation}
or equivalently
	\begin{equation}
		d\kappa  \wedge *d\kappa  = \frac{1}{{2!}}{\left( {d\kappa } \right)_{mn}}{\left( {d\kappa } \right)^{mn}}{\Omega _g} = 8{\Omega _g},
		\label{volume}
	\end{equation}
where $\Omega_g$ is the invariant volume form associated with $g$.

	A very important identity valid for any contact metric structure is \cite{Blair}
	\begin{equation}
		{\Omega _g} = \frac{{{{\left( { - 1} \right)}^n}}}{{{2^n}n!}}\kappa  \wedge {\left( {d\kappa } \right)^n}.
	\end{equation}

	Let us focus on  $n = 2$, namely a 5-dimensional contact manifold. Then we have
	\begin{equation}
		{\Omega _g} = \frac{1}{8}\kappa  \wedge d\kappa  \wedge d\kappa. 
	\end{equation}

	Now that $d\kappa$ is a horizontal 2-form, we can decompose it according to self-duality
	\begin{equation}
		d\kappa  = d{\kappa _ + } + d{\kappa _ - }
	\end{equation}
Then, one the one hand, using (\ref{vanishing-product})
	\begin{equation}
		  8{\Omega _g} = \kappa  \wedge d{\kappa _ + } \wedge d{\kappa _ + } + \kappa  \wedge d{\kappa _ - } \wedge d{\kappa _ - },
	\end{equation}
while on the other hand, following from (\ref{volume}),
	\begin{equation}
		8{\Omega _g} = d\kappa  \wedge *d\kappa  = \kappa  \wedge d{\kappa _ + } \wedge d{\kappa _ + } - \kappa  \wedge d{\kappa _ - } \wedge d{\kappa _ - },
	\end{equation}
which together with the other equation above implie
	\begin{equation}
		d{\kappa _ - } = 0,
	\end{equation}
and $d\kappa$ is naturally a self-dual 2-form.

	To summarize, for any contact structure on a smooth 5-manifold $M$, it is natural and always possible to associate a set of quantity $(\kappa, R, g)$, such that everywhere on $M$
	\begin{equation}
		\kappa \left( R \right) = 1,\;\;g\left( {R, \cdot } \right) = \kappa \left(  \cdot  \right),\;\;{\iota _R}d\kappa  = 0,\;\;{\iota _R}*d\kappa  =  d\kappa ,\;\;d\kappa  \wedge *d\kappa  = 2{\Omega _g}.
	\end{equation}
This set of conditions are exactly those arising from the $Q$-complex (\ref{Q-transformation}) and invariance of action (\ref{Lagrangian}) when $\lambda \ne 0$, namely the second row of (\ref{conditions}). This justifies our statement that the partition function $Z$ and other observables discussed earlier are homotopy invariants of contact structures.

	It is an interesting and important fact that homotopic contact structures are actually equivalent as contact structures on compact contact manifolds. A theorem by Gray  states that if $\kappa_t$ is a smooth family of contact 1-forms, then there exists a family of diffeomorphisms $f_t : M \to M$, such that $f_0 = {\rm id}$ and $f_t^* \kappa_t = \lambda_t \kappa_0$, where $\lambda_t$ is nowhere-vanishing function for any $t$. That means $\kappa_t$ defines contact structures equivalent to $\kappa_0$, namely they only differ by a diffeomorphism. Therefore, the homotopy invariants we had are actually invariants of contact structures (up to equivalence) on contact manifold $M$.
	
	Note that on a given smooth contact manifold, there may be many inequivalent contact structures. Inequivalent contact structures on simply-connected 5-manifolds have been found\cite{Koert}. More insterestingly, it has been shown that $S^5$, $T^2 \times S^3$ and $T^5$ carry infinitely many inequivalent contact structures\cite{Ustilovsky}\cite{Bourgeois}. 

	\vspace{20pt}

	When $(\kappa, R, g, \Phi)$ form a contact metric structure, there are interesting differential relations between the quantities. In the rest of this section we will demonstrate some formula that will be used in later sections.
	
	Denote the Levi-civita connection associated to $g$ as $\dot \nabla$. Then
	\begin{equation}
		\dot \nabla _n \left( {R^m \kappa _m } \right) = ( {\dot \nabla _n R^m } )\kappa _m  + R^m \left( {\nabla _n \kappa _m } \right) = 0 \Rightarrow \kappa _m \dot\nabla _n R^m  = R^m \dot\nabla _n \kappa _m  = 0,
	\end{equation}
and immediately
	\begin{equation}
		R^m \left( {d\kappa } \right)_{mn}  = R^m \dot \nabla _m \kappa _n  - R^m \dot \nabla _n \kappa _m  = R^m \dot \nabla _m \kappa _n  = 0\Rightarrow R^m \dot \nabla_m R^n = 0,
	\end{equation}
namely Reeb vector field $R$ is geodesic.

	Without proof, we point out that another important property of contact metric structure is
	\begin{equation}
		R^m \dot \nabla _m {\Phi ^n} _k  = 0.
	\end{equation}

	All the above combined give yet another formula:
	\begin{equation}
		\dot  \nabla _m R^n  =  -{ \Phi ^n} _m - \frac{1}{2} {\left( {\Phi \circ \mathcal{L}_R \Phi } \right)^n} _m,
	\end{equation}
where the tensor ${\left( {\Phi \circ \mathcal{L}_R \Phi } \right)^n} _m  = {{\Phi ^n} _k \left( {\mathcal{L}_R \Phi } \right)^k} _m $  satisfies anticommutivity
	\begin{equation}
		{\left( {\Phi \circ \mathcal{L}_R \Phi } \right)^n} _m  =  - {\left( {\mathcal{L}_R \Phi \circ \Phi}  \right)^n}_m.
	\end{equation}

	An important fact is that, for a contact metric structure, $\mathcal{L}_R \Phi = 0$ if and only if $\mathcal{L}_R g = 0$. Let us call a contact metric structure $(\kappa, R, g, \Phi)$ a K-contact structure, if $R$ is a Killing vector field with respect to the associated metric $g$, where the letter ``K" stands for "Killing". On such geometry,
	\begin{equation}
		{\dot \nabla _m}{R^n} =  - {\Phi ^n}_m.
		\label{K-contact-eq}
	\end{equation}

	\vspace{20pt}
	\underline{Examples}
	
	To end this section, let us discuss a few simple examples of contact structures.
	
	One simp example would be the Hopf-fibration $\mathbb{C}P^n = S^{2n+1}/U(1)$. One consider
	\begin{equation}
		{S^{2n + 1}} = \left\{ {\left( {{z_i}} \right) \in {\mathbb{C}^{n + 1}}|\sum\limits_i {{{\left| {{z_i}} \right|}^2}}  = 1} \right\}.
	\end{equation}
The $U(1)$ action on $\mathbb{C}^{n+1}$ sending ${z_i} \to {e^{i\theta }}{z_i}$ descends to a free $U(1)$-action on $S^{2n+1}$. Define a 1-form $\kappa$ on $\mathbb{C}$
	\begin{equation}
		\kappa  \equiv \operatorname{Im} \sum\limits_i {{{\bar z}_i}d{z_i}} ,
	\end{equation}
and also denote its restriction on $S^{2n+1}$ as $\kappa$, then it is easy to show that $\kappa$ is a contact structure on $S^{2n+1}$. The contact metric structure associated consists of: the round metric $g$, $(1,1)$-tensor $\Phi$ induced from the complex structure $J $ on $\mathbb{C}^{n+1}$, and the Reeb vector field generates the $U(1)$-action.

	More generally, given any symplectic manifold $(M_4, \omega)$ such that $\omega \in H^2(M,\mathbb{Z})$, one can form a principal $U(1)$-bundle $\pi_{M_4} : M \to M_4$, such that the connection on the bundle space $M$ is $\kappa$ and $\omega$ pulls back to be the curvature $d\kappa$. One can see that $M$ is a contact manifold with $\kappa$ being the contact 1-form. One can choose an associated contact metric structure which consists of the pull back of a associated metric on $M_4$, and the $(1,1)$-tensor comes from an almost complex structure on $M_4$ that is compatible with $\omega$.
	
	Interestingly, one can show that\cite{Blair}, any compact regular contact manifold $M$ is actually a principal $U(1)$-bundle over some symplectic manifold $M_4$ of integral type\footnote{Note that for a given compact manifold $M$ that admits regular contact structure, there could be many inequivalent ways to view it as a $U(1)$-principal bundle over symplectic 4-manifold of integral type, and therefore, $M$ admits many inequivalent regular contact structures\cite{Hamilton:2010aa}}. Moreover, since the contact metric structure on $M$ can be chosen to be the pull back of data on $M_4$, we can form a K-contact structure out of the regular contact structure. In this sense, a regular contact structure is a K-contact structure. However, a K-contact structure does not need to be regular, and there are a huge number of known examples of irregular K-contact structures.

	\section{Canonical ${\rm Spin}^\mathbb{C}$ Structure and Dirac Operator}{\label{Appendix-C}}
	
	In this appendix we will focus on 5-dimensional contact metric manifold, with contact 1-form $\kappa$, Reeb vector field $R$ and compatible metric $g$.
	
	Let us denote the vector bundle $T_HM$ of horizontal tangent vectors. Then the 2-form $d\kappa$ restricted on $T_H M$ is a symplectic form on the vector bundle. Moreover, the tensor $\Phi$ defines an almost complex structure on $T_HM$, and therefore the complexified bndle $T_HM^\mathbb{C}$ can be decomposed as
	\begin{equation}
	    {T_H M_{\mathbb{C}} = T_H^{1,0} M \oplus T_H^{0,1}M}.
	\end{equation}
Similarly we have the dual
    \begin{equation}
         {T_H M^*_\mathbb{C} = T_H^{1,0} M^* \oplus T_H^{0,1}M^*},
    \end{equation}
and one can obtain the bundle of horizontal forms $\wedge^{p,q}T_H M^*$ by taking exterior product

    Let us consider a pair of orthogonal basis $\{E_a, a = 1,2,3,4\}$ and $\{E^a, a=1,2,3,4\}$ for $T_H M$ and $T_HM^*$. We also define complex basis $\{e_i, \bar{e}^i, i = 1,2\}$ and $\{e^i, \bar{e}^i, i = 1,2\}$ for $T_HM_\mathbb{C}$ and $T_H M^*_\mathbb{C}$. The real and complex basis is related by
    \begin{equation}
        E^1 + iE^2 = e^1,\;\;E^3+ i E^4 = e^2,
    \end{equation}
and
    \begin{equation}
       \frac{1}{2}\left( E_1 - i E_2 \right)= e_1,\;\;\frac{1}{2} \left(E_3 - iE_4 \right)= e_2.
    \end{equation}
    Now let us denote $S \equiv \wedge^{0,\bullet}T_HM^*$. We also define an action of $T_H M$ on $S$ by
    \begin{equation}
        E \cdot \psi \equiv \sqrt{2}\left( E_{\bar i} {\bar e}^i \wedge \psi - g^{{\bar i}j}E_j \iota_{e_{\bar i}} \psi \right), \;\; \forall E = E_i e^i + E_{\bar i} {\bar e}^i \in \Omega^1_H(M), \psi \in \Gamma(S).
    \end{equation}
	
	It is easy to check that the action is Clifford. Namely
	\begin{equation}
	   \{ E \cdot, F\cdot\} \psi = - 2 g(E,F) \psi., \;\; \forall E, F \in \Omega^1_H(M).
	\end{equation}
Such an action can be straightforwardly extend to $\Omega^1_H(M)_\mathbb{C}$ as well as the whole exterior algebra through equation
    \begin{equation}
        (E^a \wedge E^b \wedge ...) \cdot \psi \equiv E^a\cdot E^b \cdot ... \cdot \psi.,\;\;a \ne b  \ne ...
    \end{equation}
Therefore, the bundle $S$ is actually a canonical ${\rm Spin}^\mathbb{C}$ bundle over the contact metric manifold $M$.

    One can also define a chiral operator $\Gamma_5\psi= - (E^1\wedge...\wedge E^4)\cdot \psi$. $S$ is decomposed according to chirality
    \begin{equation}
	    S = S_+ \oplus S_-,\;\; S_+ = T_HM^* \oplus T^{0,2}_HM^*, \;\;S_- =  T^{0,1}_H M^*.
	\end{equation}
	
	Choosing a basis for $S_+$ and $S_-$ as
	\begin{equation}
		\left\{ {1,\;\;\frac{1}{2}\bar e^1  \wedge \bar e^2 } \right\} {\;\;\rm and\;\;} \{\frac{1}{\sqrt{2}}{\bar e}^1, \frac{1}{\sqrt{2}}{\bar e}^2\},
	\end{equation}
one can identify the 1-forms $E^a$ with their action on $S_\pm$ as $4\times 4$ matrices:
	\begin{equation}
		\begin{gathered}
		  E^1  = \left( {\begin{array}{*{20}c}
		   0 & { - \sigma ^3 }  \\
		   {\sigma ^3 } & 0  \\
		 \end{array} } \right)\;\;\;\;\;\;E^2  = \left( {\begin{array}{*{20}c}
		   0 & {iI_{2 \times 2} }  \\
		   {iI_{2 \times 2} } & 0  \\
		 \end{array} } \right) \hfill \\
		  E^3  = \left( {\begin{array}{*{20}c}
		   0 & { - \sigma ^1 }  \\
		   {\sigma ^1 } & 0  \\
		 \end{array} } \right)\;\;\;\;\;\;E^4  = \left( {\begin{array}{*{20}c}
		   0 & { - \sigma ^2 }  \\
		   {\sigma ^2 } & 0  \\
		 \end{array} } \right) \hfill \\ 
		\end{gathered}
	\end{equation}
form which one conclude the isomorphism $T_HM^*_\mathbb{C} = Hom(S_+, S_-) = Hom(S_-, S_+)$. Similarly one can identify the horizontal 2-forms $\wedge_\pm^2 T_HM^*_\mathbb{C} = End_0 (S_\pm)$, where we $\wedge_\pm^2 T_H M^*$ is the bundle of (anti)self-dual horizontal 2-forms, and $End_0$ denotes the traceless endomorphisms. For instance, we have for $E^1\wedge E^2 + E^3\wedge E^4 \in \Omega_H^+(M)$
	\begin{equation}
		E^1  \wedge E^2  + E^3  \wedge E^4  =  - 2i\left( {\begin{array}{*{20}c}
		   {\sigma ^3 } & 0  \\
		   0 & 0  \\
		 \end{array} } \right)
	\end{equation}
which maps $\psi \in S_+$ to $S_+$. The trace part of the $End(S_\pm)$ corresponds to multiplication of a function and therefore we have isomorphism
	\begin{equation}
		End\left( {S_ \pm  } \right) = S_ \pm   \otimes S_ \pm ^*  =  \wedge _ \pm ^2 T_H M_\mathbb{C}^*  \oplus   \mathbb{C},
		\label{isomorphism_1}
	\end{equation}
where $\mathbb{C}$ denote a trivial comoplex line bundle generated by complex function or complex-valued top form. Actually we have also isomorphisms:
	\begin{equation}
		T_H M_\mathbb{C}^*  = End\left( {S_ \pm  ,S_ \mp  } \right) = S_ \pm   \otimes S_ \mp ^* ,
		\label{isomorphism_2}
	\end{equation}
	\begin{equation}
		\wedge^\bullet T_HM^*_\mathbb{C} = End(S,S) = S\otimes S^*
		\label{isomorphism_3}.
	\end{equation}

	We can further twist the any of the spinor bundles with some vector bundle $V$. One obtain new Clifford module $S\otimes V$ with Clifford multiplication defined as
	\begin{equation}
		E \cdot \left( {\psi  \otimes \sigma _V } \right) \equiv \left( {E \cdot \psi } \right) \otimes \sigma _V .
	\end{equation}
One important example of such construction is when $V = S^*$, and we have the canonical isomorphism (\ref{isomorphism_3}) to bundle of complex-valued forms mentioned above. The induced Clifford multiplication is
	\begin{equation}
		E^a  \cdot \psi  = E^a  \wedge \psi  - \iota \left( {E_a } \right)\psi .
	\end{equation}
Similarly we can take $V = S_+^*$, and the Clifford multiplication mapping $S_- \otimes S_+^* \to S_+ \otimes S_+^*$, or equivalently $T_H M_\mathbb{C}^*  \to  \wedge _ + ^2 T_H M_\mathbb{C}^*  \oplus \mathbb{C}$ is
	\begin{equation}
		E^a  \cdot \eta  = \pi _H^ + (  E^a  \wedge \eta ) - \iota \left( {E^a } \right)\eta.
		\label{Clifford_1}
	\end{equation}

	One can now define Dirac-like operator on the spinor bundle $S$. One canonical choice is to use the generalized Tanaka-Webster connection. On any contact metric manifold there exists a cononical metric connection $\nabla$ on $TM$ that satisfies
	\begin{equation}
		\nabla \kappa = \nabla R = \nabla g = 0.
		\label{GTW_1}
	\end{equation}
Since it is a metric connection, its connection coefficient can be written in terms of Christoffel symbol and contorsion tensor $C$:
	\begin{equation}
		{\Gamma^k}_{mn} = {\dot \Gamma^k}\;_{mn} + {C^k}_{mn},
	\end{equation}
where $\dot \Gamma$ denotes the Levi-civita connection, and the contorsion tensor reads\footnote{The coordinate-free definition is \cite{Tanno}
	\begin{equation}
		  {\nabla _X}Y = {{\dot \nabla }_X}Y + \kappa \left( X \right)\Phi (Y) - \kappa \left( Y \right){{\dot \nabla }_X}R + \left( {{{\dot \nabla }_X}\kappa } \right)\left( Y \right)R.
	\end{equation}
	}
	\begin{equation}
		\begin{gathered}
  		{C^k}_{mn} = \kappa _m {\Phi ^k }_n  + \kappa _n ( {\Phi  + \frac{1}{2}\Phi \mathcal{L}_R \Phi } )^k \;_m  +\frac{1}{2} R^k \left( {d\kappa } \right)_{mn}  \hfill \\
  		\;\;\;\;\;\;\;\;\;\;\;\;\; + \frac{1}{4}R^k {\left( {\mathcal{L}_R \Phi } \right)^p }_m \left( {d\kappa } \right)_{pn}  \hfill .\\ 
		\end{gathered} 
	\end{equation}

	Because $\nabla$ preserves $\kappa$ and therefore the contact distribution $T_H M$, it naturally induces a connection on $T_H M$ and therefore the spinor bundle $S$, which we still denote as $\nabla$. With such connection, we can define a Dirac-like operator $\slashed{D} : \Gamma(S) \to \Gamma(S)$ as
	\begin{equation}
		\slashed{D} \psi \equiv \sum\limits_{a = 1}^4 {E^a  \cdot \nabla _{E_a } \psi } ,\;\; \forall \psi\in \Gamma(S).
		\label{Kohn-Dirac_1}
	\end{equation}
as well as its restriction on $\slashed{D}^\pm : \Gamma(S_\pm ) \to \Gamma(S_\mp)$. The operator $\slashed{D}$ is called Kohn-Dirac operator in mathematical literatures \cite{Petit2005229}.
	
	Now let us twist the $S_\pm$ by vector bundle $V = S_+^*$, and the resulting operator $\slashed{D}^-_{S_+^*}$ reads
	\begin{equation}
		\slashed{D}_{S_ + ^* }^ -  \psi  = \sum\limits_a {E^a  \cdot \nabla _{E_a } \psi } = \sum\limits_a {\pi _H^ +  \left( {E^a  \wedge \nabla _{E_a } \psi } \right) - \iota \left( {E_a } \right)\nabla _{E_a } \psi }  ,\;\; \forall \psi \in \Omega_H^1(M)_\mathbb{C},
		\label{Kohn-Dirac_2}
	\end{equation}
where we used the isomorphism (\ref{isomorphism_2}) and (\ref{Clifford_1}).

	Note that the above construction originates from the canonical ${\rm Spin^\mathbb{C}}$-bundle over symplectic manifolds $(M, \omega)$, which can be defined with some compatible almost complex structure. Dirac operator can also be defined by some metric connection that also preserves the almost complex structure.
	
	Let us end this appendix by mentioning a few interesting properties of the generalized Tanaka-Webster connection $\nabla$. As we have shown before, Although $\nabla$ was originally a connection on $TM$, it is by itself also a connection on $E \equiv T_H M$. The tensor $\Phi$ defines an almost complex structure on $E$, and we can view $E$ as a complex vector bundle with connection $\nabla$ over $M$ associated to the contact metric structure. Let us consider the case where this structure is actually a K-contact structure. Let $\sigma$ be a section of $E$ (which is essentially some horizontal vector field), then we have
	\begin{equation}
		{\nabla _R}\sigma  = {\dot \nabla _R}\sigma  + \Phi (\sigma ) = {\dot \nabla _R}\sigma  - {\dot \nabla _\sigma }R = \mathcal{L}_R \sigma.
	\end{equation}
where (\ref{K-contact-eq}) has been used. The above condition implies that the parallel transport of the connection $\nabla$ along Reeb integral curves coincides with the Reeb flow. In particular, let $\gamma$ be a closed integral curve based at $p$, then the holonomy
	\begin{equation}
		ho{l_\nabla }\left( {\gamma ,p} \right) = {\Psi _{\gamma ,p}},
	\end{equation}
where $\Psi_{\gamma, p}$ is the linearized return map: $E_p \to E_p$. Note that this map is used to define various index in contact homology \cite{Contact-Homology}.

	\end{appendices}

\bibliographystyle{JHEP}
\bibliography{Contact_Instanton}

\end{document}